\documentclass[11pt,a4paper]{amsart}
\usepackage[utf8]{inputenc}
\usepackage{amsmath}
\usepackage{amsfonts}
\usepackage{amssymb}
\usepackage{graphicx}
\usepackage{subfigure}
\usepackage{booktabs}
\usepackage{tikz}
\usetikzlibrary{decorations.pathreplacing,calc}
\usepackage[ruled,vlined]{algorithm2e}
\usepackage[round,authoryear]{natbib}
\usepackage{bibunits}
\usepackage{float}

\usepackage{nicefrac}
\usepackage{tabularx}
\usepackage{blindtext}

\usepackage[left=2.2cm,right=2.2cm,top=2cm,bottom=2cm]{geometry}

\newcommand{\D}{{\mathop{}\!\mathrm{d}}} 
\newcommand{\R}{\mathbb{R}}
\newcommand{\Q}{\mathbb{Q}}

\newcommand{\G}{\mathbb{G}}

\newcommand{\N}{\mathbb{N}}

\newcommand{\PP}{\mathbb{P}}

\newcommand{\E}{\mathbb{E}}

\numberwithin{equation}{section}  
\newtheorem{defn}{Definition}[section]
\newtheorem{exa}[defn]{Example}
\newtheorem{rem}[defn]{Remark}

\usepackage{hyperref}
\usepackage{xcolor}
\hypersetup{
    colorlinks,
    linkcolor={red!50!black},
    citecolor={blue!50!black},
    urlcolor={blue!80!black}
}

\title[Empirical Analysis of the Model-Free Valuation Approach]{Empirical Analysis of the Model-Free Valuation Approach: Hedging Gaps, Conservatism, and Trading Opportunities}
\author[Zixing Chen, Yihan Qi, Shanlan Que, 
Julian Sester, 
Xiao Zhang]{Zixing Chen$^{1}$, Yihan Qi$^{1}$, Shanlan Que$^{1}$, 
Julian Sester$^{1}$, 
Xiao Zhang$^{1}$}

\begin{document}

\maketitle

\begin{center}
\normalsize{\today} \\ \vspace{0.5cm}
\small\textit{
$^{1}$National University of Singapore, Department of Mathematics,\\ 21 Lower Kent Ridge Road, 119077.}                                                                                                     
                        
\end{center}

\begin{abstract}~In this paper we study the quality of model-free valuation approaches for financial derivatives by systematically evaluating the difference between model-free super-hedging strategies and the realized payoff of financial derivatives using historical option prices from several constituents of the S$\&$P~$500$ between $2018$ and $2022$.

Our study allows us in particular to describe the realized gap between payoff and model-free hedging strategy empirically so that we can quantify to which degree model-free approaches are overly conservative. Our results imply that the model-free hedging approach is only marginally more conservative than industry-standard models such as the Heston model while being model-free at the same time.

This finding, its statistical description and the model-independence of the hedging approach enable us to construct an explicit trading strategy which, as we demonstrate, can be profitably applied in financial markets, and additionally possesses the desirable feature of explicit control of its downside risk due to its model-free construction preventing losses pathwise.
\end{abstract}
\section{Introduction}

{ Robust and model-free pricing approaches date back at least to the seminal work by Hobson (\cite{hobson1998robust}) who showed that lookback and related path-dependent options admit tight, model-independent price bounds and explicit super-replicating strategies based solely on observed call prices and the martingale property, without assuming any specific asset dynamics.}
{ More than a decade later, \cite{cox2011robust2} and \cite{cox2011robust} established model-free upper and lower bounds on the arbitrage-free price of different types of barrier options purely from observed vanilla option prices by constructing explicit super- and sub-hedging strategies and showing these bounds are tight via Skorokhod embedding techniques, all without assuming a specific probabilistic model for the underlying asset’s dynamics.}

{ Based on \cite{cox2011robust}, in \cite{obloj2010robust} the authors analyzed how robust hedging strategies for digital double barrier options perform compared to traditional model-specific hedges (such as delta or delta/vega hedging). They find that under model uncertainty and transaction costs, robust hedges often achieve similar or better risk-adjusted performance and lower downside risk than classical methods across a range of market scenarios, albeit evaluated in hypothetical model-based forward markets.}

{ In the 2010s, extending the above mentioned pathwise and Skorokhod-based pricing approaches, the literature on model-free derivative valuation developed rapidly. Building} on the foundational works \cite{beiglbock2013model, beiglbock2016problem, henry2016explicit}, which establish the connection between solutions of the martingale optimal transport problem, model-free price bounds for financial derivatives and associated hedging strategies, the literature on model-free approaches and their application to derivative valuation has grown rapidly and still continues to grow. See, for instance, \cite{ansari2024improved, cheridito2017duality, chiu2023model, davis2016model, dolinsky2014martingale, de2020bounds, fahim2016model, hobson2012robust, hobson2015robust, eckstein2021robust, lux2017improved}, among many others.
Using only available market data of liquidly traded put- and call options, the aforementioned model-free approaches compute price bounds for less liquidly traded (so called exotic) options solely under the paradigm of no arbitrage and of consistency with observed liquid option prices but, importantly, without using any assumptions on the stochastic dynamics of the underlying assets. The resultant price bounds are then calculated as extreme prices among all calibrated arbitrage-free models, and therefore can be understood to be model-free and to respect the so called \emph{Knightian uncertainty} (\cite{knight1921risk}), i.e., the unquantifiable risk of having chosen a wrong model.

While appealing from a risk management perspective, not imposing model assumptions theoretically comes  at the application-prohibitive cost of resultant model-free price  bounds by construction being larger than the price in any calibrated model that is also consistent with liquid option prices and arbitrage-free, making the approach less appealing from a practitioner's perspective, see also the discussions in \cite{eckstein2021martingale,lutkebohmert2019tightening,dolinsky2018super}.

{ However, despite the extensive theoretical development of model-free pricing and the comparative performance analysis of robust hedges, apart from \cite{obloj2010robust}, it remains largely unexplored how conservative model-free price bounds and the corresponding hedging strategies are when evaluated against realized payoffs in real financial markets.}

According to the duality theory developed in \cite{beiglbock2013model, beiglbock2016problem, henry2016explicit}, to study the quality of model-free upper price bounds, one can equivalently study prices of model-free super-replication strategies (or sub-replication strategies for lower price bounds).

{ Hence, empirical analysis of realized super-hedging payoffs provides direct information on the effective tightness of model-free price bounds in practice.
Namely, if the gap between the realized payoff of a price-bound-attaining model-free super-replication strategy and the realized derivative payoff is persistently large across many observed price paths, this indicates that the strategy’s initial price, and hence its initial cash endowment, is overly conservative. In such cases, the initial capital could be reduced to obtain a less conservative strategy whose realized payoff is closer to the derivative’s actual payoff, with high probability under the empirical distribution induced by the observed paths.

This phenomenon is intrinsic to model-free hedging: super-replication strategies are constructed to dominate the derivative payoff pathwise, i.e., for every admissible price path. Since many of these admissible paths are highly unrealistic and effectively never occur in practice, the resulting hedges can appear excessively expensive when evaluated under empirically observed distributions.}

Relying on the above outlined idea, the present paper aims at closing an existing research gap by studying systematically the difference between model-free hedging strategies (corresponding to model-free price bounds) and realized payoffs of exotic derivatives. For our empirical study, we use historical option prices written on seven of the largest assets by market capitalization of the S$\&$P~$500$ between $2018$ and $2022$. This paper is, to the best of our knowledge, one of the first systematic empirical studies quantifying the realized hedging performance and conservatism of model-free super-hedging strategies using real market data. Our findings challenge the prevailing common belief that model-free methods are impractically conservative, at least if the involved assets and sufficiently many call- and put options are liquidly traded in the market, as it is the case for our study.

Our analysis shows that model-free hedging strategies lead to a hedging performance comparable to that of industry-standard models such as the Black–Scholes, Heston, and Dupire local volatility models. As, in contrast to the aforementioned industry-standard models, model-free super-hedging strategies do not rely on a specific probabilistic model of the underlying security, they always lead to larger payoffs than the derivative given any realization of the security. { From a practitioner’s perspective, this analysis suggests that model-free bounds may be sufficiently tight to support economically meaningful trading decisions.} We use this robustness property to propose a trading strategy that makes use of both our statistical description of the payoff gap and the model-free construction of the hedging strategy. Trading for this strategy is triggered whenever observed market quotes for the derivative are close enough to its model-free price bounds where the degree of proximity determines the probability of a positive payoff which can be exactly described according to our statistical analysis. The strategy is robust in the sense that up to the initial investment no additional money can be lost due to its model-free construction, consequently leading to a low-risk trading strategy with an easy-to-understand risk profile.

{ While \cite{obloj2010robust} study the performance of robust hedging strategies relative to classical hedges in hypothetical model-based markets, our paper provides, to the best of our knowledge, the first systematic empirical analysis of the conservativeness of model-free price bounds themselves using real market data, and leverages this insight to construct a provably low-risk trading strategy.}

The remainder of this paper is as follows. In Section~\ref{sec:Data_and_tech} we provide an overview of the data used, Section~\ref{sec:methodology} describes in detail the employed mathematical approach to compute the super-hedging strategies based on linear programming. Section~\ref{sec:results} presents the results of our analysis when being applied to data, and Section~\ref{sec:tradingstrat} discusses and analyzes a trading strategy that is based on our statistical findings.

\section{Data and technology}\label{sec:Data_and_tech}
\subsection{Data}\label{sec:data}
For our study, we  use data received from \emph{Wharton Research Data Services}. The data comprises bid and ask prices of traded put- and call options in the time between Jan 1 2018 and Dec 31 2022. The considered underlying securities on which the options are written are the stocks of Amazon.com Inc, JP Morgan Chase \& Co., Johnson \& Johnson, Procter \& Gamble Co., Walmart Inc, VISA Inc, and Tesla Inc.

\subsection{Code}\label{sec:code}
For the sake of transparency and replicability, we provide  the used Python-code in an accompanying GitHub-repository\footnote{The repository can be found under \href{https://github.com/QIYIHAN/Empirical-Price-Bounds}{https://github.com/QIYIHAN/Empirical-Price-Bounds}.}. Unfortunately, legal constraints prevent us from providing the historical option price data in the GitHub repository.

\section{Methodology}\label{sec:methodology}
Our methodology consists mainly of three steps. In the first step we pre-process the data, second we compute model-free price bounds and corresponding hedging strategies via linear programming, and third we compute the gap between the realized payoff of an exotic financial derivative and the payoff of the previously computed hedging strategy given outcomes of the underlying assets described in Section~\ref{sec:data}. In the following we introduce the mathematical setup in Section~\ref{sec:model-free-valuation} and detail all the computational steps in Section~\ref{sec:lp} and \ref{sec:gap}.
\subsection{Model-free valuation of financial derivatives}\label{sec:model-free-valuation}
We consider some financial derivative providing a payoff $\Phi(S_{t_1},\dots,S_{t_n})$ at some future time $t_n>0$ with $S_t$ denoting the value of the underlying asset at time $t$.
To determine a range of model-free prices for $\Phi$, we compute the minimal super-replication price of $\Phi$ when investing in so called semi-static trading strategies (see, e.g. \cite{davis2007range} for more details on semi-static trading), i.e., trading strategies investing once in liquidly traded European call options and adjusting dynamically its investment in the underlying security. These strategies possess a payoff at time $t_n$ of the form
\begin{equation}\label{strategy_transaction_costs}
\begin{aligned}
 u_{d,(\theta_{i,j}),(H_i)}: \R^n &\rightarrow \R \\
 (S_{t_1},\dots,S_{t_n}) &\mapsto e^{r_{(t_n-t_0)}\cdot(t_n-t_0)}d \\
 &+\sum_{i=1}^n\sum_{j=1}^{M_i} e^{r_{(t_n-t_i)}\cdot(t_n-t_i)}(\theta_{i,j}^{\rm call, +}-\theta_{i,j}^{\rm call, -}) (S_{t_i}-K_{i,j}^{\rm call})^+ \\
 &{ +\sum_{i=0}^{n-1}e^{r_{(t_{n}-t_0)}\cdot (t_n-t_0)}H_i(S_{t_1},\dots,S_{t_i}) (\widetilde{S}_{t_{i+1}}-\widetilde{S}_{t_{i}})-\sum_{i=0}^{n-1}e^{r_{(t_n-t_i)}\cdot(t_n-t_i)}\varepsilon|\Delta H_i|,}
\end{aligned}
\end{equation}
where $r_{\tau}\in \R$ denotes the interest rate used for discounting of a time period of length $\tau$, and where  $\widetilde{S}_{t_{i}}:=S_{t_i}e^{-r_{(t_{i}-t_0)}\cdot(t_{i}-t_0)}$ denotes the discounted stock price. Moreover, $d \in \R$ denotes the cash position, $(\theta_{i,j}^{\rm call,+}, \theta_{i,j}^{\rm call,-}) \in \R_{\geq 0}^2$ the holdings of the call options, $K_{i,j}^{\rm call } \in \R_{\geq 0}$ the strikes of the call options, $M_i \in \N$ the number\footnote{Note that we do not need to distinguish between the number of strikes considered for the call options since we can simply use the larger of the two and include options multiple times if necessary to match the number of options considered.} of the call options with maturity $t_i$, $H_i(S_{t_1},\dots,S_{t_i}) \in \R$ (while $H_0$ is a constant) denote the amount of the investment in the underlying security at time $t_i$. The expression $\varepsilon|\Delta H_i|$ with $|\Delta H_i|:= |H_i(S_{t_1},\dots,S_{t_i})-H_{i-1}(S_{t_1},\dots,S_{t_{i-1}})|$ is accounting for per-share transaction costs $\varepsilon>0$ incurred from changing the position $H_i$, where $H_{-1} \equiv 0$. In our empirical implementation, we use $\varepsilon = 0.0035$ USD per share\footnote{This choice of $\varepsilon$ is implied by \cite{interactivebrokers}. According to \cite{interactivebrokers} the per share transaction costs are not larger than $0.0035\$$}. Transaction costs are assumed to be paid at the rebalancing date $t_i$. Therefore, in undiscounted maturity-$t_n$ units, the transaction cost incurred at $t_i$ is multiplied by $e^{r_{(t_n-t_i)}\cdot(t_n-t_i)}$.

\begin{rem}
\begin{itemize}

\item[(i)] For the sake of presentation we have decided to not include put options into \eqref{strategy_transaction_costs} which is well justified by the put-call parity (\cite{stoll1969relationship}) and since the call options we consider are sufficiently liquid by our data selection process described in Section~\ref{sec:preprocess}.
\item[(ii)] Note that the trading dates $(t_i)_{i=1,\dots,n}$ of the dynamic strategy in \eqref{strategy_transaction_costs} coincide exactly with the expiration dates of the considered call options. This is a natural construction, as allowing for more trading dates beyond the expiration dates, does not  change the resultant price bounds computed in \eqref{eq:super_replication}, see also \cite[Corollary 2.1]{henry2017model} and \cite[Lemma 4.2]{sester2023intermediate}.
\end{itemize}
\end{rem}

To compute the price of the strategy with payoff $ u_{d,(\theta_{i,j}),(H_i)}$, we denote the ask and bid prices of the considered call options by $h_{i,j}^{\rm call,+}$ and $h_{i,j}^{\rm call,-}$ respectively. Then, to compute the largest model-free price for $\Phi$ which does not allow for model-free arbitrage (in line with the arbitrage notion from \cite[Definition 1.2]{acciaio2016model}), we compute the super-hedging price by
\begin{equation}
\begin{aligned}\label{eq:super_replication}
\overline{\operatorname{P}}(\Phi):= \inf_{d,(\theta_{i,j}),(H_i)}\Bigg\{&d+\sum_{i=1}^n\sum_{j=1}^{M_i}(\theta_{i,j}^{\rm call, +}h_{i,j}^{\rm call, +}-\theta_{i,j}^{\rm call, -}h_{i,j}^{\rm call, -}) ~ \\
&\text{ subject to}~u_{d,(\theta_{i,j}),(H_i)}(S) \geq \Phi(S) \text{ for all } S \in \mathbb{R}_{\geq 0}^n \Bigg\}
\end{aligned}
\end{equation}
compare also the expositions in the related literature, e.g., in \cite{acciaio2016model}, \cite[Section A.1]{neufeld2021model} or \cite[Section II]{neufeld2022deep}. Correspondingly, a sub-hedging price can be computed via 
\begin{equation}\label{eq:sub_replication}
\underline{\operatorname{P}}(\Phi):=-\overline{\operatorname{P}}(-\Phi).
\end{equation}

By construction, market prices outside the interval $[\underline{\operatorname{P}}(\Phi),\overline{\operatorname{P}}(\Phi)]$ allow one to pocket a model-free arbitrage profit through investing in the corresponding semi-static sub- or super-hedging strategies with price attaining the bounds of $[\underline{\operatorname{P}}(\Phi),\overline{\operatorname{P}}(\Phi)]$ - independent of the realized outcome of the underlying security.
{
\begin{rem}\label{rem:discounting}
Throughout the paper, the target payoff $\Phi(S_{t_1},\dots,S_{t_n})$ is specified in
\emph{undiscounted} currency units at maturity $t_n$. For this reason,
the superhedging condition
\[
u_{d,(\theta_{i,j}),(H_i)}(S_{t_1},\dots,S_{t_n}) \;\ge\; \Phi(S_{t_1},\dots,S_{t_n})
\]
is formulated in undiscounted terms, so that both sides of the inequality are expressed
in the same units. This choice is without loss of generality. Introducing the deterministic discount factors
$B(t)=e^{r_{(t-t_0)}(t-t_0)}$ and the discounted asset $\tilde S_t:=S_t/B(t)$, the above
inequality is equivalent to the discounted condition
\[
\tilde u(\tilde S)\;\ge\;\tilde\Phi(\tilde S),
\qquad
\tilde u:=\frac{u}{B(t_n)}, \quad \tilde\Phi:=\frac{\Phi}{B(t_n)}.
\]
Hence, the discounted and undiscounted formulations differ only by a deterministic change
of num\'eraire. For clarity, we may equivalently write the discounted strategy payoff as
\begin{align*}
\tilde u_{d,(\theta_{i,j}),(H_i)}
&= d
+ \sum_{i=1}^n\sum_{j=1}^{M_i}
(\theta^{\mathrm{call},+}_{i,j}-\theta^{\mathrm{call},-}_{i,j})
\cdot e^{-r_{(t_i-t_0)}(t_i-t_0)} (S_{t_i}-K^{\mathrm{call}}_{i,j})_+\,  \\
&\qquad  + \sum_{i=0}^{n-1}
\left(H_i(S_{t_1},\dots,S_{t_i})\big(\tilde S_{t_{i+1}}-\tilde S_{t_i}\big)
- e^{-r_{(t_i-t_0)}(t_i-t_0)} \varepsilon|\Delta H_i|\right).
\end{align*}
Separating the discounted time--$t_0$ costs of the static option positions yields a
discounted gain process that coincides with the notion of \emph{discounted trading gains} employed, e.g., in
\cite[Section~2 \& Section~3.1]{cheridito2017duality}. Consequently, the super-replication
functional from \eqref{strategy_transaction_costs} is exactly the standard discounted super-replication price applied to the
discounted payoff $\tilde\Phi$, rewritten in undiscounted units to facilitate direct
comparison with $\Phi$.
\end{rem}
}
\begin{rem}
Note that if a financial model  is calibrated to observed bid-ask prices of call options, then arbitrage-free prices of an exotic derivative $\Phi$ can be computed by  $\E_{\Q}\left[e^{-r_{(t_n-t_0)}(t_n-t_0)} \Phi(S_{t_1},\dots,S_{t_n})\right]$ where $\Q$ is a martingale measure that fulfils $\E_{\Q}\left[e^{-r_{(t_i-t_0)}(t_i-t_0)}(S_{t_i}-K_{i,j}^{\rm call})^+\right] \in \left[h_{i,j}^{\rm{call},-},~h_{i,j}^{\rm{call},+}\right]$ for all $i,j$. If we denote the set of all such probability measures by $\mathcal{Q}$, then \cite[Section 3.1.3]{cheridito2017duality} or \cite[Section A.1]{neufeld2021model} show that under mild assumptions
\[
\overline{\operatorname{P}}(\Phi) = \sup_{\Q\in \mathcal{Q}} \E_{\Q}\left[e^{-r_{(t_n-t_0)}(t_n-t_0)} \Phi(S_{t_1},\dots,S_{t_n})\right],
\]
i.e., the smallest super-replication price $\overline{\operatorname{P}}(\Phi)$ can also be understood as the largest price among all arbitrage-free financial models calibrated to market data.
\end{rem}

\subsection{Computation of model-free super-hedges with linear programming}\label{sec:lp}
For the computation of model-free super-hedging price $\overline{\operatorname{P}}(\Phi)$ as defined in \eqref{eq:super_replication} and the associated super-hedging strategy we mainly rely on a numerical routine using linear programming (LP) as  outlined in similar settings in \cite[Section 3]{henry2013automated} and \cite{guo2019computational}. Since the computation of the sub-hedging strategy follows analogously through using the relation in \eqref{eq:sub_replication}, we focus with our presentation on the computation of a super-hedging strategy at minimal price
$\overline{\operatorname{P}}(\Phi)$ in the two time-step case $n=2$, i.e.,
\begin{equation}
\begin{aligned}\label{eq:super_replication_twotimestepcase}
\overline{\operatorname{P}}(\Phi):= \inf_{d,(\theta_{i,j}),H_0,H_1}\Bigg\{&d+\sum_{i=1}^2\sum_{j=1}^{M_i}(\theta_{i,j}^{\rm call, +}h_{i,j}^{\rm call, +}-\theta_{i,j}^{\rm call, -}h_{i,j}^{\rm call, -}) \\
&\text{ subject to}~u_{d,(\theta_{i,j}),(H_0,H_1)}(S) \geq \Phi(S) \text{ for all } S \in \mathbb{R}_{\geq 0}^2 \Bigg\}.
\end{aligned}
\end{equation}
The above problem \eqref{eq:super_replication_twotimestepcase} falls in the class of infinite-dimensional linear programming problems. To solve \eqref{eq:super_replication_twotimestepcase} numerically, we aim at transforming \eqref{eq:super_replication_twotimestepcase} to a finite dimensional problem. To this end, we first consider a (sufficiently large) grid
\[
\G=(\G^1,\G^2) \subset \R_{\geq 0}^2,
\]
where both $\G^1$ and $\G^2$ are one-dimensional grids with $N$ meshes that discretize the values attained by the distributions of $S_{t_1}$ and $S_{t_2}$, respectively. Given a function $H_1: \R \rightarrow \R$, representing the dynamic investing in the underlying security, as defined in \eqref{strategy_transaction_costs}, for each grid point $S_{t_1}^\ell \in\G^1$, we define the value $H_1^\ell:=H_1(S_{t_1}^\ell)$. Finally, to approximate $H_1$ on the whole reals, we apply
a cubic spline interpolation (see, e.g., \cite{ahlberg2016theory}) using $(H_1^\ell)_{\ell=1,\dots,N}$, and obtain a cubic spline function $$\rm{spline}\left((H_1^\ell)_{\ell=1,\dots,N}\right): \R \rightarrow \R.
$$
which coincides with $H_1$ on $\mathbb{G}^1$ and interpolates smoothly the non-grid values. The above outlined procedure allows us to define an approximating finite dimensional linear programming problem\footnote{Note that for the formulation in \eqref{eq:super_replication_grid}, we do not necessarily need a smooth function approximation as only the grid points on $\mathbb{G}$ are considered for the super-replication condition. However, in later sections we will use the corresponding cubic-spline function to represent the learned strategy also applied to values off the grid.}
\begin{equation}
\begin{aligned}\label{eq:super_replication_grid}
\widetilde{\overline{\operatorname{P}}}(\Phi,\G):= \inf_{d,(\theta_{i,j}),H_0,(H_1^\ell)_{\ell}}\Bigg\{&d+\sum_{i=1}^2\sum_{j=1}^{M_i}(\theta_{i,j}^{\rm call, +}h_{i,j}^{\rm call, +}-\theta_{i,j}^{\rm call, -}h_{i,j}^{\rm call, -}) \\~
&\text{ subject to }~u_{d,(\theta_{i,j}),(H_0,\rm{spline}\left((H_1^\ell)_{\ell}\right))}(S) \geq \Phi(S) \text{ for all } S \in \G\Bigg\}.
\end{aligned}
\end{equation}
To find numerical solutions to $\widetilde{\overline{\operatorname{P}}}(\Phi,\G)$, we apply a cutting plane algorithm as outlined in \cite[Section 3.1]{henry2013automated} of which we provide details in the appendix, compare Algorithm~\ref{alg:cutting_plane}.

\subsection{Computation of the gap between the realized payoff and the payoff of the hedging strategy}\label{sec:gap}
This section outlines the method employed to compute the difference between the outcomes of minimal model-free super-replication strategies (as well as maximal  model-free sub-replication strategies) and the underlying derivative payoff.
\subsubsection{Preprocessing of the data}\label{sec:preprocess}
From the option price data described in Section~\ref{sec:data} we first select all pairs of option prices with the same underlying stock and fixed length of maturity difference $t_2-t_1 > 0$, where $t_1, t_2$ are two future times relevant for the option payoff. To obtain sufficiently reliable prices, for each fixed maturity difference, we select the $20$ most liquid call options (and associated strikes) by choosing those with the highest trading volume. Then, to avoid arbitrage among the considered options (in particular butterfly arbitrage, see \cite{gatheral2014arbitrage}), we apply the arbitrage repair method introduced in \cite{Cohen2020} to the mid-prices. Using the \emph{repaired} mid price $C^{*}$ and original spread $\delta = C_{\rm ask}-C_{\rm bid}$, we obtain eventually repaired bid and ask prices $C_{\rm bid}^{*}$ and $C_{\rm ask}^{*}$ via the relations
\begin{align*}
  C_{\rm ask}^{*} = C^{*} + \frac{1}{2}\delta,   \qquad C_{\rm bid}^{*} = C^{*} - \frac{1}{2}\delta.
\end{align*}
Further details of the preprocessing procedure are described in the accompanying GitHub repository which can be found under  \href{https://github.com/QIYIHAN/Empirical-Price-Bounds}{https://github.com/QIYIHAN/Empirical-Price-Bounds}.
\subsubsection{Computing interest rates}\label{sec:ir}
To model the interest rate $(r_t)_t$ we use throughout our study a Nelson--Siegel model (\cite{Nelson-Seigel}) fitted to U.S. treasury yields  with the historical parameters that can be found under \url{https://www.federalreserve.gov/data/nominal-yield-curve.htm}.

In the empirical implementation, interest rates are treated deterministically. For each initiation date $t_0$, we evaluate the Nelson–Siegel zero-coupon yield curve fitted to the U.S. Treasury yields observed on that date, and use the resulting deterministic discount factors for all future cash flows of the corresponding hedge. Thus, the rate $r_{\tau}$ used for maturity $\tau$ is fixed at $t_0$; we do not model stochastic interest rates or update the discount curve along the realized stock path. Given the relatively short maturities considered in our empirical study, we regard this simple deterministic discounting convention as sufficient for our purposes.

\subsubsection{Computation of the gap}
Upon computing  $\widetilde{\overline{\operatorname{P}}}(\Phi,\G)$ defined in \eqref{eq:super_replication_grid}, by using Algorithm~\ref{alg:cutting_plane}, we obtain the optimal values of the parameters $d$, $(\theta_{i,j})_{\substack{i=1,2\\ j =1,\dots,M_1 }}$, $H_0$ and $(H_1^\ell)_{\ell=1,\dots,N}$, and interpolate $(H_1^\ell)_{\ell=1,\dots,N}$ by a cubic spline interpolator $\widetilde{H}_1:=\rm{spline}\left((H_1^\ell)_{\ell=1,\dots,N}\right)$. Denoting the realized prices of the underlying security at times $t_1$ and $t_2$, respectively by $S\in \R^2$,we compute the raw value of the gap between realized hedging strategy and realized derivative payoff by 
\[
u_{d,(\theta_{i,j}),(H_0,\widetilde{H}_1)}(S)-\Phi(S).
\]

Eventually, to make the gap values comparable across different underlying securities, we normalize the raw gap by the price of the underlying security at the initial time $t_0$ denoted by $S_0$, i.e., we define the \emph{normalized} hedging gap (or simply gap in the following) by
\begin{equation}
\begin{aligned}\label{eq:gap_computation}
\operatorname{gap}:=\frac{u_{d,(\theta_{i,j}),(H_0,\widetilde{H}_1)}(S)-\Phi(S)}{S_0}.
\end{aligned}
\end{equation}

\section{Empirical Tightness of Model-Free Hedging}\label{sec:results}
In this section we provide and discuss the results of the presented numerical routine from Section~\ref{sec:methodology} applied to the data presented in Section~\ref{sec:Data_and_tech}. All experiments in this section are based on historical option quotes and realized underlying stock prices from the data described in Section~\ref{sec:data}. No simulated stock paths or simulated exotic-option market prices are used in Section~\ref{sec:results}. For each observation, $t_0$ is a historical option-quote date, $t_1$ and $t_2$
are option-expiration dates with the specified maturity difference, and $S_{t_1}$ and $S_{t_2}$ are the realized stock prices on those dates. 

\subsection{The payoff gap among different time-to-maturities}\label{sec:superhedge_forward}
\begin{figure}[h!]
    \centering
    \includegraphics[width=.9\linewidth]{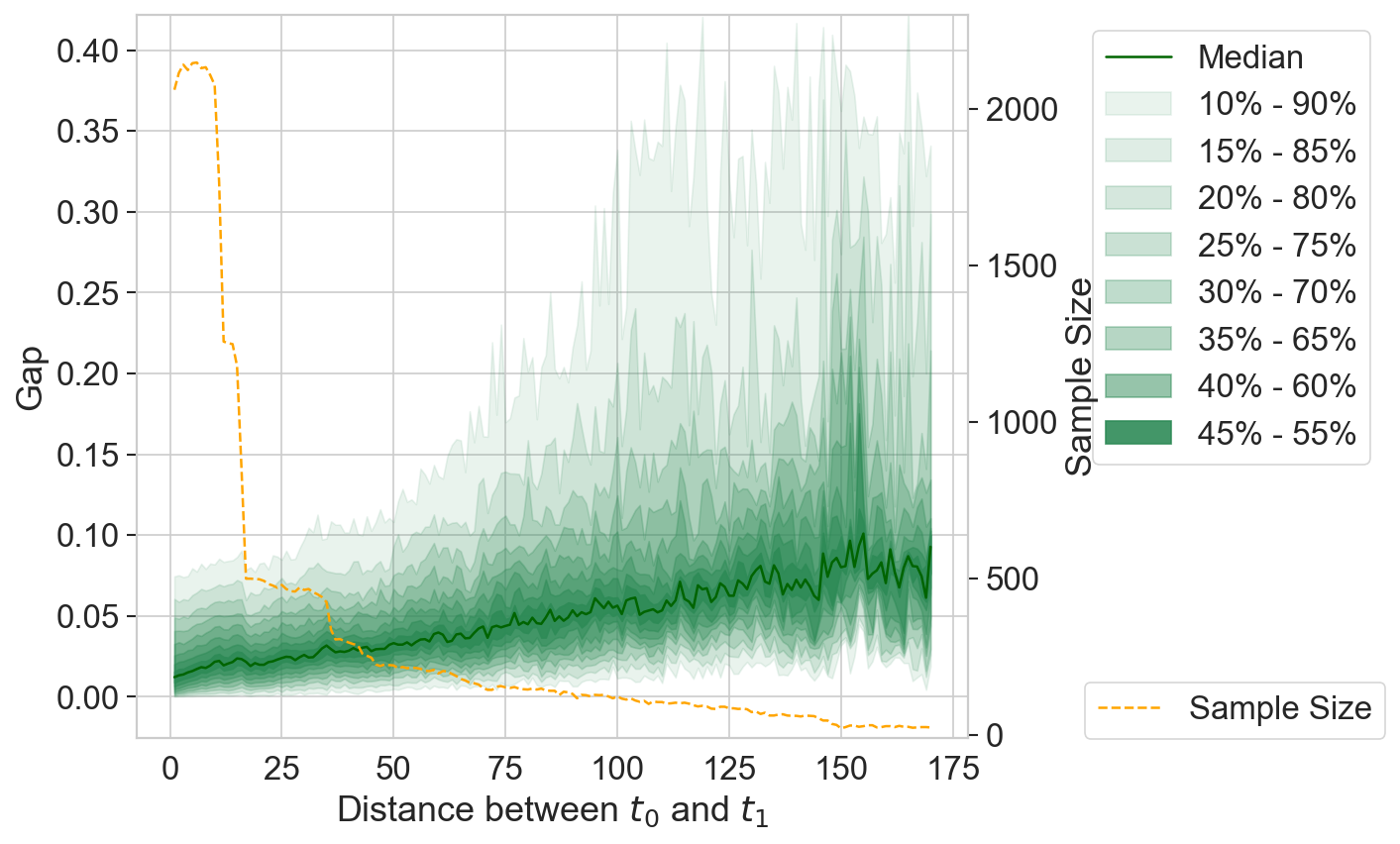}
    \caption{\textcolor{blue}{} The different quantile ranges of the payoff gap for super-hedging with payoff function $(S_{t_2}-S_{t_1})^+$.}
    \label{fig:payoff1_super}
\end{figure}

We start our analysis by considering a \emph{forward start at-the-money call option} possessing a payoff of the form  $\Phi(S_{t_1},S_{t_2}):=(S_{t_2}-S_{t_1})^+:=\max\{S_{t_2}-S_{t_1},0\}$, where $S_{t_1}$ and $S_{t_2}$ denote the value of the underlying asset at future times $t_1$ and $t_2$, respectively, i.e., this is a call option where the strike is determined at time $t_1$ such that the option is \emph{at-the-money}. Using the data described in Section~\ref{sec:data}, we evaluate the  gaps, computed according to \eqref{eq:gap_computation}, between the realized payoff of the minimal model-free super-hedging strategy and that of the forward start option across different time-to-maturities, i.e., for different distances between $t_0$ (the time when the option contract is set up) and $t_1$, while the difference $t_2-t_1$ is fixed to $28$ days. We illustrate median and different quantile ranges of the computed gaps as well as the underlying sample size in Figure \ref{fig:payoff1_super}.

We observe that with the initial date $t_0$ approaching the date $t_1$, i.e., the distance between $t_0$ and $t_1$ decreasing, the gaps tend to become smaller: the median of the payoff gap is decreasing, and the quantile ranges are shifting towards smaller gap values. Similarly, the variance of the gap becomes smaller for a decreasing time-to-maturity corresponding to the empirical fact that the farther the maturity lies in the future the higher the associated uncertainty of the outcome of the asset and the payoff (measured in volatility). These observations become more unstable with an increasing time-to-maturity, which we attribute mainly to a smaller available sample size for longer maturities, indicated by the dashed orange line in Figure \ref{fig:payoff1_super}.

\begin{figure}[h!]
    \centering
    \includegraphics[width=.9\linewidth]{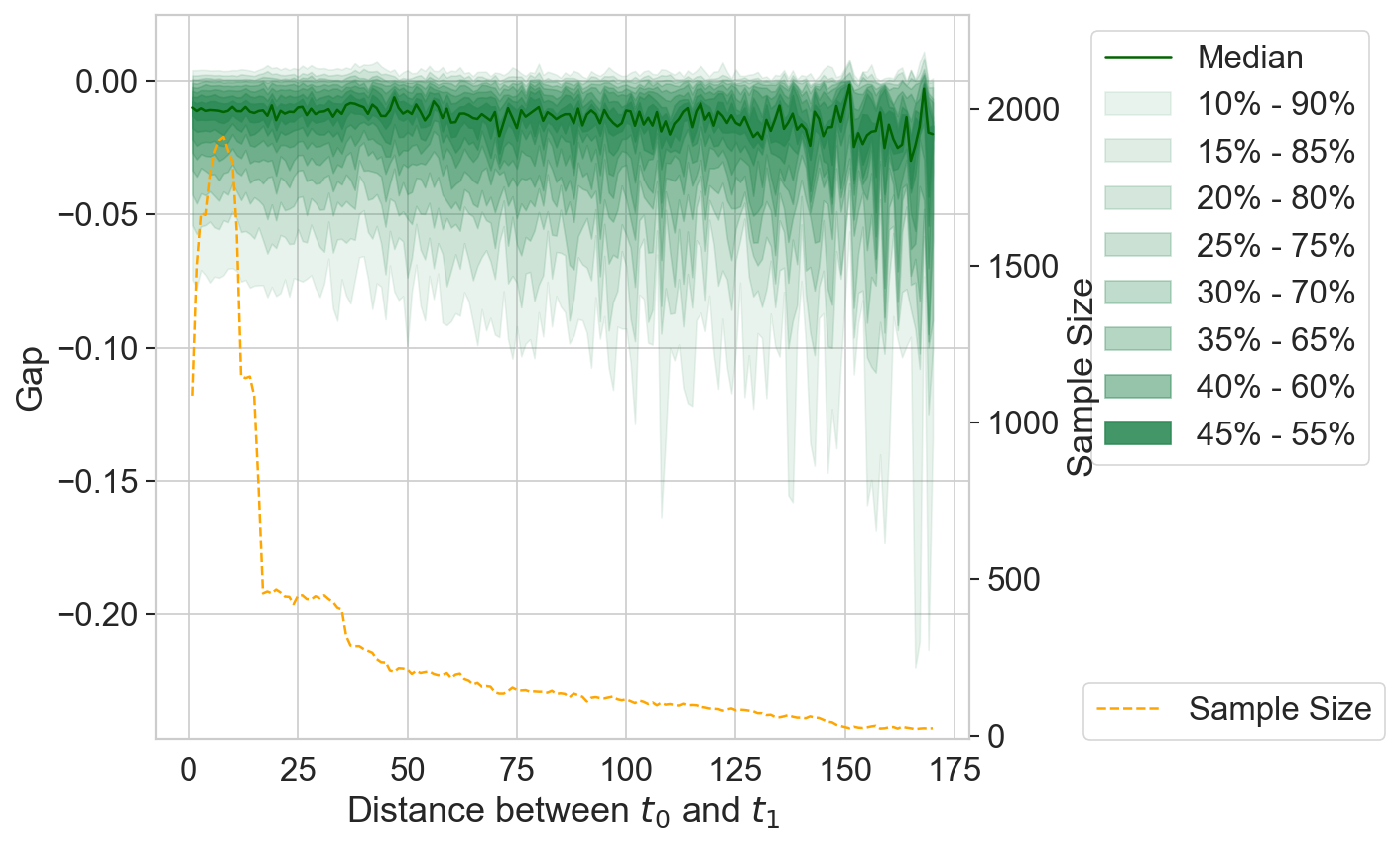}
    \caption{\textcolor{blue}{} The different quantile ranges of payoff gap for sub-hedging with payoff function $(S_{t_2}-S_{t_1})^+$.}
    \label{fig:payoff1_sub}
\end{figure}

\subsubsection{Sub-hedging}\label{sec:subhedge_forward}
In addition to the case of super-hedging discussed above, we also study the gap between realized outcomes of model-free sub-hedging strategies and the realized payoff of a forward start option with payoff $(S_{t_2}-S_{t_1})^+$. The statistics (median, quantile ranges) of the gap values under sub-hedging are illustrated in Figure \ref{fig:payoff1_sub}. We remark that according to \eqref{eq:gap_computation} the gap values in this situation are by construction mostly negative, but, in contrast to intuition, there are some mildly positive gap values in Figure \ref{fig:payoff1_sub} which we explain  through the discretization inherent in Algorithm \ref{alg:cutting_plane} in which we accept the result as long as the gap values do not exceed the tolerance level ${\rm TOL}$.

The observations in Figure \ref{fig:payoff1_sub} are mostly in line with those from Figure~\ref{fig:payoff1_super}, although the median of the gap appears to be  more stable. Overall, as in Figure~\ref{fig:payoff1_super} the empirical payoff gap widens with an increasing time-to-maturity, where the gap for the model-free sub-hedging strategy widens less significantly than the model-free super-hedging strategy in this case.

\subsection{The payoff gap among different payoffs}\label{sec:different_payoffs}
\begin{figure}[h!]
    \centering
    \includegraphics[width=1.0\linewidth]{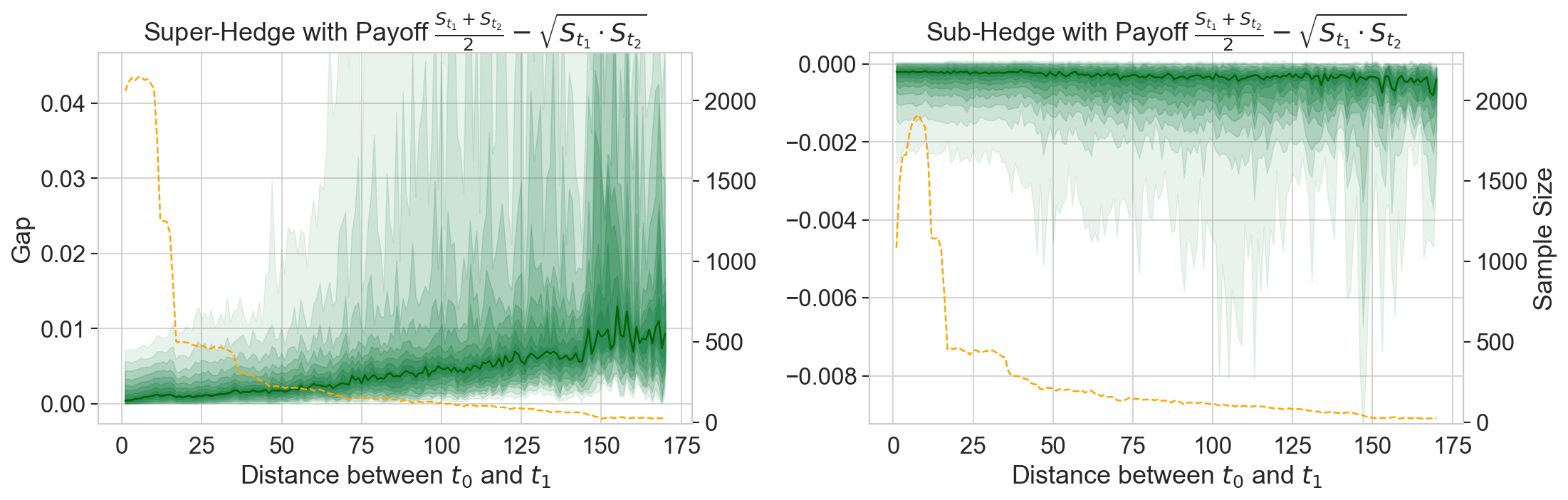}
    \caption{\textcolor{blue}{}The different quantile ranges of the payoff gap for super-hedging with payoff function $\frac{S_{t_1}+S_{t_2}}{2}-\sqrt{S_{t_1}\cdot S_{t_2}}$.}
    \label{fig:payoff7}
\end{figure}

\begin{figure}[h!]
    \centering
    \includegraphics[width=1.0\linewidth]{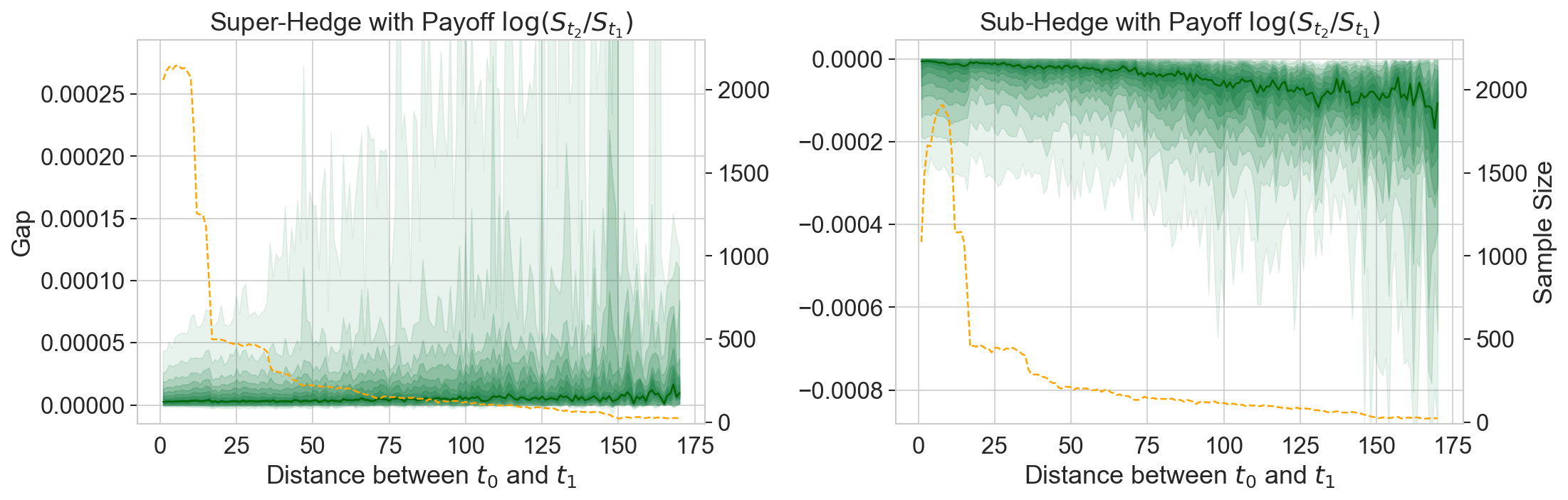}
    \caption{\textcolor{blue}{}The different quantile ranges of the payoff gap for super-hedging with payoff function $\log(S_{t_2}/S_{t_1})$.}
    \label{fig:payoff3}
\end{figure}

\begin{figure}[h!]
    \centering
    \includegraphics[width=1.0\linewidth]{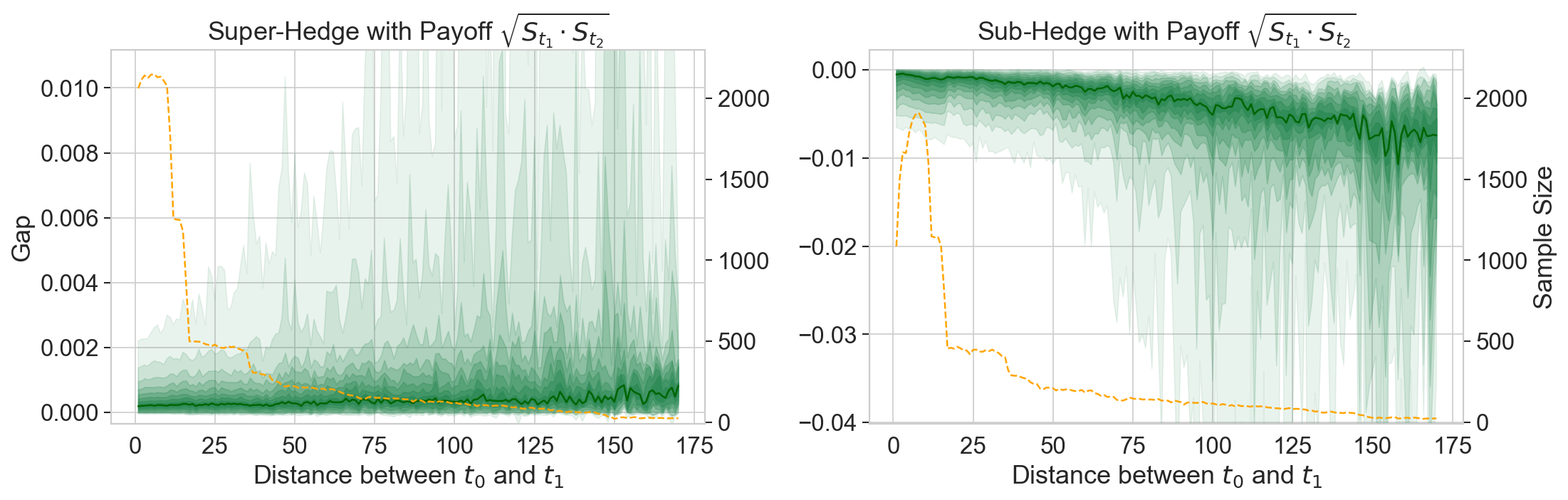}
    \caption{\textcolor{blue}{}The different quantile ranges of the payoff gap for super-hedging with payoff function $\sqrt{S_{t_1}\cdot S_{t_2}}$.}
    \label{fig:payoff4}
\end{figure}

We repeat the experiments from Section~\ref{sec:superhedge_forward}  with several other payoff functions. The corresponding statistics, shown in Figures~\ref{fig:payoff7}--\ref{fig:payoff4}, are consistent with the findings reported in Section~\ref{sec:superhedge_forward}.


Moreover, in the appendix, in Figures \ref{fig:payoff1_hist} - \ref{fig:payoff4_hist},  we illustrate the distribution of the gap values for all payoffs under both super-hedging and sub-hedging by depicting the respective histograms.

\subsection{Comparison with model-based hedges}\label{sec:model_based}
Next, to set the size of the payoff gaps reported in Section~\ref{sec:superhedge_forward} and Section~\ref{sec:different_payoffs} in relation, we compare the quality of the computed model-free super- and sub-hedging payoffs with hedging payoffs computed under several industry-standard models that are calibrated to the option data presented in Section~\ref{sec:data}. To keep the paper self-contained, we present the  considered benchmark models briefly in Section~\ref{sec:models}. The corresponding hedging strategies in each of the models are implemented by considering a delta-hedging strategy with adjustment to the respective Gamma and Theta sensitivities, compare for more details, e.g., \cite{hull2017optimal}. As the results from Section~\ref{sec:superhedge_forward} have revealed, the gap distribution of model-free hedges under different derivatives turns out to be similar. Hence, for the analysis in this section we focus solely on the forward-start call option with payoff $(S_{t_2}-S_{t_1})^+$. From initiation date $t_0$ to $t_2$, we compute the option delta on a daily basis, compare Section~\ref{sec:models} for more details.

\begin{figure}[h!]
    \centering
    \begin{minipage}[t]{0.6\textwidth}
        \centering
        \includegraphics[width=\textwidth]{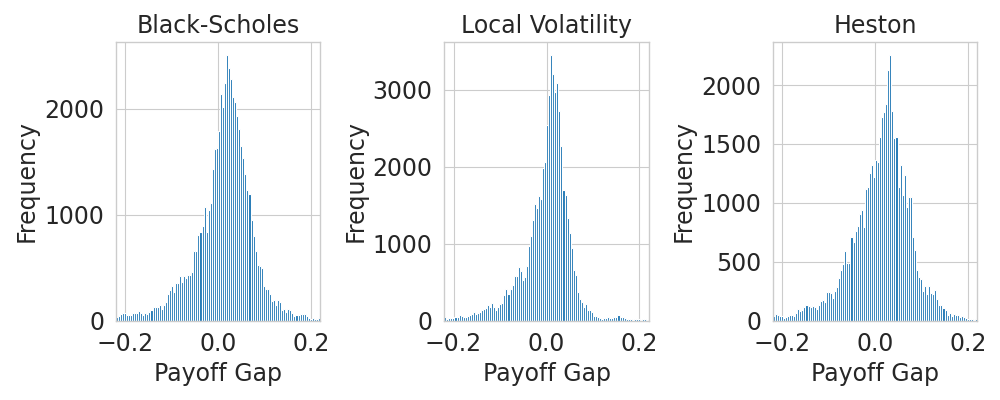}
    \end{minipage}%
    \hfill
    \begin{minipage}[t]{0.4\textwidth}
        \centering
        \includegraphics[width=\textwidth]{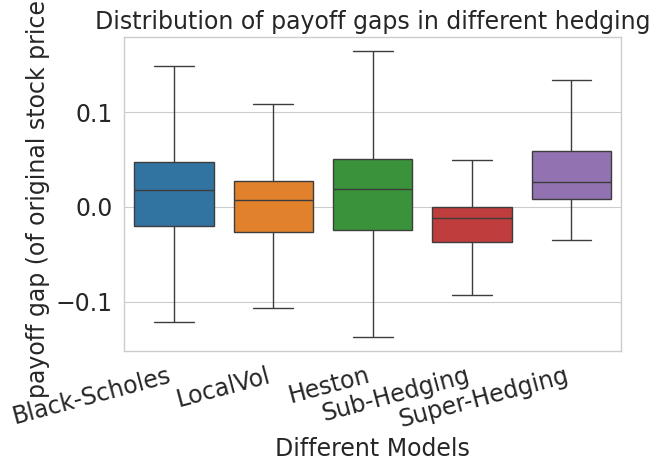}
    \end{minipage}
    \caption{\textcolor{blue}{} The distribution of the gap between derivative payoff and different hedging strategies for a derivative payoff of the form $(S_{t_2}-S_{t_1})^+$ in model-based and model-free hedging approaches.}
    \label{fig:CompareModels_box}
\end{figure}

\begin{table}[htbp]
\centering
\caption{Summary Statistics for different hedging strategies for a derivative payoff of the form $(S_{t_2}-S_{t_1})^+$ using either model-based or model-free hedging approaches.}
\label{tab:summary_stats}
\begin{tabularx}{\textwidth}{l *{5}{>{\centering\arraybackslash}X}}
\toprule
 &{\small \textbf{Black-Scholes}} & \textbf{\small Local Volatility} & \textbf{\small Heston} & \textbf{\small Super-Hedging} & \textbf{\small Sub-Hedging} \\
\midrule
Mean     & 0.00 & -0.01 & 0.02 & {0.05} & {-0.03} \\
Std.dev.      & 0.10 & 0.09 & 1.37 & {0.11} & {0.08} \\
Min      & -1.47 & -1.47 & -1.32 & {-0.03} & {-2.66} \\
Max      & 0.38 & 0.36 & 171.98 & {5.06} & {0.16} \\
25 \%      & -0.02 & -0.03 & -0.02 & {0.01} & {-0.04} \\
75 \%    & 0.05 & 0.03 & 0.05 & {0.06} & {-0.00} \\
\bottomrule
\end{tabularx}
\end{table}

In Figure~\ref{fig:CompareModels_box} and Table~\ref{tab:summary_stats}, we compare the distribution of the resultant hedging errors (defined as in \eqref{eq:gap_computation}) of the model-based hedging strategies with the hedging errors of the model-free super and sub-hedging strategies among all derivatives whose payoffs are $(S_{t_2}-S_{t_1})^+$  where the difference between $t_1$ and $t_2$  is fixed to $28$ days, and where the hedges are evaluated on the data described in Section~\ref{sec:data}. We observe that the hedges under all  considered models lead to reasonably tight hedging gaps with interquartile ranges of $[-0.02,0.05], [-0.03,0.03]$, and $[-0.02,0.05]$, respectively. The main observation is though that the interquartile ranges of the gaps of model-free super-hedges and sub-hedges are given by $[0.01,0.06]$, and $-[0.04,0.00]$, respectively and therefore they are of similar size as the model-based hedges, contradicting the intuition and common belief that model-free approaches lead to overly conservative hedging outcomes. As by pricing-hedging duality (see, e.g., \cite{eberlein2008duality}, \cite{hou2018robust}), prices of hedges correspond to arbitrage-free prices of the corresponding derivative, the prices produced by model-free pricing approaches turn out to be competitive and rather close to prices produced by industry standard models (and thus traded prices of OTC derivatives) - an observation that we will use in the next section to explicitly construct  a profitable trading strategy.

\section{A trading strategy based on model-free price bounds}\label{sec:tradingstrat}

The analysis from Section~\ref{sec:results} allows us to derive a trading strategy with an easy-to-understand risk control regarding its downside risk. The idea behind the strategy is to exploit discrepancies between market prices and model-free super- or sub-hedging price bounds: the study from Section~\ref{sec:model_based} shows that price bounds of model-free approaches are relatively close to model-based prices of industry standard models; if now indeed the observed price of a derivative lies \emph{sufficiently close} to these bounds, a model-free hedge is constructed whose payoff dominates the derivative, thereby allowing the investor to lock in a non-negative payoff exceeding the limited initial investment with a high probability. By tuning a threshold parameter (that we later call $c$), the investor can control the trade-off between risk  and expected return.
\subsection{Description of the strategy}\label{sec:description_strat}
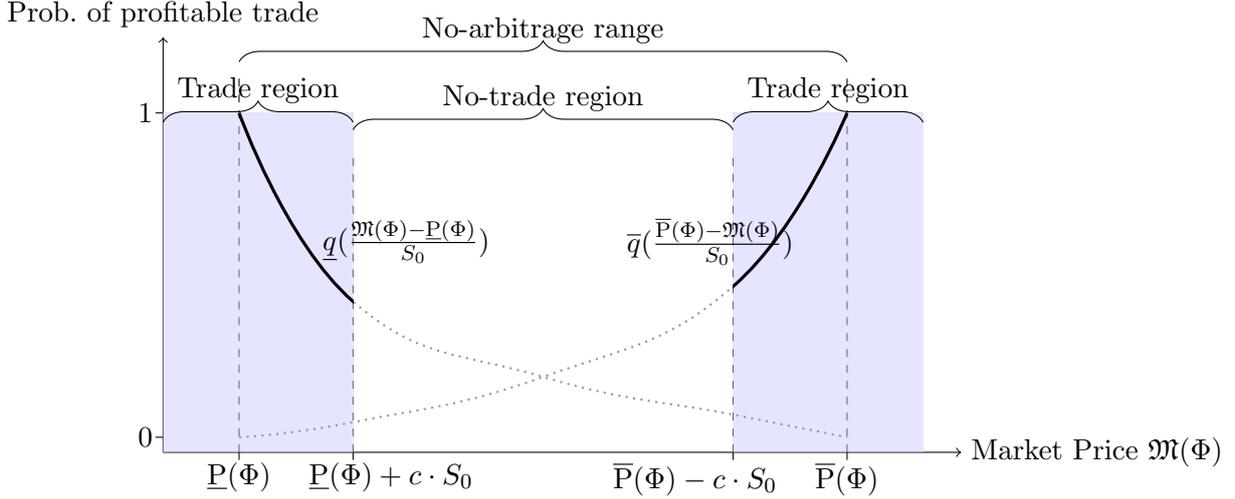
\begin{figure}
    \centering

\begin{tikzpicture}[scale=1]
  \draw[->] (0,0) -- (10.5,0) node[right] {Market Price ${\mathfrak{M}}(\Phi)$};
  \draw[->] (0,0) -- (0,5.5) node[above] {Prob.\ of profitable trade};

  \draw (0,0.2) -- (-0.1,0.2);
  \draw (0,4.5) -- (-0.1,4.5);

  \draw (1,0) -- (1,-0.1);
  \draw (2.5,0) -- (2.5,-0.1);
  \draw (7.5,0) -- (7.5,-0.1);
  \draw (9,0) -- (9,-0.1);
  
  \draw[dashed] (1,0) -- (1,5);
  \draw[dashed] (2.5,0) -- (2.5,4);
  \draw[dashed] (7.5,0) -- (7.5,4);
  \draw[dashed] (9,0) -- (9,5);

  \node[left, align=left] at (0,0.2) {$0$};
  \node[left, align=left] at (0,4.5) {$1$};
  \node[below, align=left] at (1,0) {$\underline{\mathrm{P}}(\Phi)$};
  \node[below, align=right] at (3,0) {$\underline{\mathrm{P}}(\Phi)+c\cdot S_0$};
  \node[below, align=center] at (7,0) {$\overline{\mathrm{P}}(\Phi)-c\cdot S_0$};
  \node[below] at (9,0) {$\overline{\mathrm{P}}(\Phi)$};

  \fill[blue!20, opacity=0.5] (0,0) -- (0,4.5) -- (2.5,4.5) -- (2.5,0) -- cycle;
  \fill[blue!20, opacity=0.5] (7.5,0) -- (7.5,4.5) -- (10,4.5) -- (10,0) -- cycle;

  \draw[thick, black!40, dotted] plot[smooth, tension=0.8] coordinates {(1,4.5) (2.5,2) (5,1) (7.5,0.5) (9,0.2)};
  \draw[thick, black!40, dotted] plot[smooth, tension=0.8] coordinates {(1,0.2) (2.5,0.4) (5,1.0) (7.5,2.2) (9,4.5)};

  \begin{scope}
    \clip (0,0) rectangle (2.5,6);
    \draw[very thick] plot[smooth, tension=0.8] coordinates {(1,4.5) (2.5,2) (5,1) (7.5,0.5) (9,0.2)};
  \end{scope}
  \begin{scope}
    \clip (7.5,0) rectangle (10,6);
    \draw[very thick] plot[smooth, tension=0.8] coordinates {(1,0.2) (2.5,0.4) (5,1.0) (7.5,2.2) (9,4.5)};
  \end{scope}

  \node at (3.2,2.8) {$\underline{q}(\tfrac{{\mathfrak{M}}(\Phi)-\underline{\rm{P}}(\Phi)}{S_0})$};
  \node at (7.2,2.8) {$\overline{q}(\tfrac{\overline{\rm{P}}(\Phi)-{\mathfrak{M}}(\Phi)}{S_0})$};

  \draw [decorate, decoration={brace, amplitude=8pt}, yshift=5pt]
    (1,5) -- (9,5) node [midway, yshift=12pt] {No-arbitrage range};
  \draw [decorate, decoration={brace, amplitude=8pt}, yshift=5pt]
    (0,4.2) -- (2.5,4.2) node [midway, yshift=12pt] {Trade region};
  \draw [decorate, decoration={brace, amplitude=8pt}, yshift=5pt]
    (2.5,4.1) -- (7.5,4.1) node [midway, yshift=12pt] {No-trade region};
  \draw [decorate, decoration={brace, amplitude=8pt}, yshift=5pt]
    (7.5,4.2) -- (10,4.2) node [midway, yshift=12pt] {Trade region};
\end{tikzpicture}

    \caption{The figure illustrates the idea of the trading strategy: the strategy is initiated whenever the market price ${\mathfrak{M}}(\Phi)$ is \emph{close enough} to one of the price bounds $\overline{\mathrm{P}}(\Phi)$, $\underline{\mathrm{P}}(\Phi)$, illustrated via the blue shaded area. The probabilities $\underline{q}(\tfrac{{\mathfrak{M}}(\Phi)-\underline{\rm{P}}(\Phi)}{S_0})$ and $\overline{q}(\tfrac{\overline{\rm{P}}(\Phi)-{\mathfrak{M}}(\Phi)}{S_0})$ describe the probability for a profitable trade in dependence of the market price.}
    \label{fig:prof_trade}
\end{figure}

To construct the strategy, let $\overline{u_{d,(\theta_{i,j}),(H_i)}}$ denote a hedging strategy of the form described in \eqref{strategy_transaction_costs} attaining the upper price bound $\overline{\operatorname{P}}(\Phi)$\footnote{We can always find such a strategy up to an arbitrarily small tolerance $\varepsilon>0$ which we ignore here for our exposition.}. Then, we have by construction of the model-free super-hedge that
$\overline{u_{d,(\theta_{i,j}),(H_i)}}(S) -\Phi(S) \geq 0$ for all $S \in \R^n$. Moreover, for all levels of relative hedging gaps $c\in \R$ there exists an associated probability $\overline{q}(c)\in [0,1]$ such that the empirical probability of a realized relative gap larger than $c$ is $\overline{q}(c)$, i.e.,
\[
\widehat{\PP}\left(\frac{\overline{u_{d,(\theta_{i,j}),(H_i)}}(S) -\Phi(S)}{S_0} \geq c \right) = \overline{q}(c),
\]
where  $\widehat{\PP}$ denotes the empirical distribution derived from historical stock data $(S^{(i)})_{i=1,\dots,N}=(S_{t_1}^{(i)},S_{t_2}^{(i)})_{i=1,\dots,N}$ of the respective stock, i.e., 
\[
\widehat{\PP}  = \frac{1}{N}\sum_{i=1}^N\delta_{(S_{t_1}^{(i)}, S_{t_2}^{(i)})},
\]
for $\delta_x$ denoting the Dirac measure at point $x\in \R^2$ and where $N$ denotes the number of historical stock prices used for the construction of $\widehat{\PP}$.
In the same manner, we denote by $\underline{u_{d,(\theta_{i,j}),(H_i)}}$ a strategy attaining the lower price bound $\underline{\operatorname{P}}(\Phi)$. Then, for all $c\in \R$ there exists some $\underline{q}(c)\in [0,1]$ such that
\[
\widehat{\PP}\left(\frac{\Phi(S)-\underline{u_{d,(\theta_{i,j}),(H_i)}}(S)}{S_0} \geq c \right) = \underline{q}(c).
\]
If now an observed market price $ \mathfrak{M}(\Phi)$ for the derivative $\Phi$ is close to either $\overline{\operatorname{P}}(\Phi)$ or $\underline{\operatorname{P}}(\Phi)$, i.e., given some $c >0$, we have 
\begin{equation}\label{eq:min_prices}
\min\{\mathfrak{M}(\Phi)-\underline{\operatorname{P}}(\Phi),~~\overline{\operatorname{P}}(\Phi)-\mathfrak{M}(\Phi)\}\leq c \cdot S_0,
\end{equation}
then we may invest in $\overline{u_{d,(\theta_{i,j}),(H_i)}}(S) -\Phi(S)$, or $\Phi(S)-\underline{u_{d,(\theta_{i,j}),(H_i)}}(S)$ (we choose whatever is the binding condition in \eqref{eq:min_prices}). For such an investment we pay less than $c \cdot S_0$ to subsequently hold a strategy which leads under $\widehat{\PP}$ with a probability of either $\overline{q}(\tfrac{\overline{\rm{P}}(\Phi)-{\mathfrak{M}}(\Phi)}{S_0}) \geq \overline{q}(c)$ or $\underline{q}(\tfrac{{\mathfrak{M}}(\Phi)-\underline{\rm{P}}(\Phi)}{S_0}) \geq \underline{q}(c)$   to a positive profit. Moreover, and this makes the investment extremely attractive and fundamentally different from other \emph{statistical arbitrage} related concepts, the maximal loss is the initial investment $c \cdot S_0$ as the held strategy upon initialization, by its construction as a model-free hedge, never leads to negative payoffs. This means an investor can fix the level $c$ which imposes an upper bound on potential future losses. We illustrate the idea in Example~\ref{exa1} and Figure~\ref{fig:prof_trade}.

\begin{exa}\label{exa1}
Consider a forward-start call option on Tesla with payoff
\[
\Phi(S_{t_1}, S_{t_2}) = (S_{t_2} - S_{t_1})^+,
\]
where $S_0 = 800$ (Tesla’s spot price at initial time $t_0$), $t_1 =$ March 1, and $t_2 =$ March 29. Further, assume that the computed model-free bounds, computed via Algorithm~\ref{alg:cutting_plane}, are given by
\[
\underline{\rm P}(\Phi) = 34 \quad \text{and} \quad \overline{\rm P}(\Phi) = 44,
\]
and that the market price of the derivative is $\mathfrak{M}(\Phi) = 42.5$, close to the upper bound. Then the trade is triggered since
\[
\min\{ \mathfrak{M}(\Phi) - \underline{\rm P}(\Phi), \overline{\rm P}(\Phi) - \mathfrak{M}(\Phi) \} = 1.5 \leq c \cdot S_0,
\]
for a threshold $c = 0.02$. We thus invest $\overline{C} = \overline{\rm P}(\Phi) - \mathfrak{M}(\Phi) = 1.5$  by selling the option and buying the model-free super-hedge, which dominates the derivative payoff path-wise. Assume that on maturity we observe $S_{t_1} = 810$ and $S_{t_2} = 870$, so that:
\[
\Phi(S_{t_1}, S_{t_2}) = (870 - 810)^+ = 60.
\]
Note that, by construction, the model-free super-hedge leads to a payoff that dominates the derivative payoff $\Phi(S_{t_1}, S_{t_2})$ by a margin of at least $1.5$ with probability $\overline{q}(c)$. If we now assume that the model-free super-hedge realizes a payoff of $\overline{u}(S_{t_1}, S_{t_2}) = 63$, then the payoff of our strategy is
\[
\overline{u}(S_{t_1}, S_{t_2})- \Phi(S_{t_1}, S_{t_2}) = 3, \quad \text{and net profit} = 3 - 1.5 = 1.5.
\]
This leads in this example to a 100\% return with a worst-case loss of at most $1.5$, highlighting the favorable risk–return profile of the strategy.
\end{exa}

The choice of $c$ allows one to steer the level of risk entailed in the strategy: namely, the larger $c$, the less restrictive is condition \eqref{eq:min_prices}, and hence more trades are initiated coming however at the cost of a reduced probability of a profitable outcome (since $\underline{q}(\cdot)$ and $\overline{q}(\cdot)$ are decreasing functions). Moreover, the possible initial price of the strategy increases with $c$, and therefore also the associated potential losses, implying an increased downside deviation of the outcomes. Finally, we remark that even though the estimation of the empirical measure $\widehat{\PP}$ introduces model risk, the maximum downturn $c \cdot S_0$ is independent of $\widehat{\PP}$ as it is based on a model-free trading strategy, i.e., since $\Phi(S)-\underline{u_{d,(\theta_{i,j}),(H_i)}}(S) \geq 0$, and $\overline{u_{d,(\theta_{i,j}),(H_i)}}(S) -\Phi(S) \geq 0$ pathwise, i.e., for all $S \in \R^n$.
\subsection{Testing the strategy}
To empirically test the strategy described in Section~\ref{sec:description_strat}, we use the data outlined in Section~\ref{sec:Data_and_tech} and split it into a training period (to construct $\widehat{\PP}$), and a testing period (to backtest the strategy). All the derivatives considered for super- and sub-hedging in Section~\ref{sec:results} are included for the empirical study of the strategy. The data related to the period from Jan $2018$ to Sep $2020$ is used as training period, the second half (Oct $2020$ to Dec $2022$) serves as test period.

Since we do not have access to historical prices of OTC-traded exotic derivatives, as a realistic surrogate we simulate derivative prices $\mathfrak{M}(\Phi)$ by a calibrated Heston model.
{ We choose the Heston model as a representative stochastic-volatility benchmark, as forward-start options are particularly sensitive to forward variance dynamics; similar conclusions are expected for other well-calibrated smile-consistent models. The Heston model parameters are calibrated on a daily basis for each underlying stock using all available call option quotes, the underlying spot price, and the corresponding interest-rate inputs observed on that day (see the data description in Section~\ref{sec:data} and interest rate calculation in Section~\ref{sec:ir}). We estimate the parameters by minimizing the squared pricing error between market option prices (mid-quotes) and model-implied prices, and we add a Tikhonov regularization term to stabilize the parameter estimates over time. We implement the calibration in \emph{Python} using \emph{QuantLib} (\cite{quantlib}), which provides an analytic European-option Heston pricer (semi-closed-form/Fourier-based). Further details of the calibration are described in Appendix \ref{subsubsec:heston}. Then, using the calibrated parameters, we simulate daily market prices via  Monte Carlo simulation, performing 50,000 simulation paths with 1,000 time steps per option. Our Heston calibration can occasionally produce option prices that violate the no-arbitrage bounds. This is not unexpected: the empirical option data itself may contain residual arbitrage, and the calibration is further subject to numerical and approximation errors. To avoid artificially improving the reported performance by trading on model-implied mispricings that are driven by these artifacts rather than genuine opportunities, we impose a conservative filter: whenever a Heston model price falls outside the no-arbitrage range, we discard that option and do not allow trading in it. This filtering step is conservative and biases results against finding arbitrage trades, since model-induced arbitrage opportunities are excluded by construction.}

{ The use of surrogate model prices at this step can be justified from a practical viewpoint, as it has been well studied that after calibration, the Heston model reproduces the broad shape of observed implied-volatility smiles/skews reasonably well—particularly away from very short maturities, see, e.g. \cite{bakshi1997empirical}, \cite{fiorentini2002estimation} and \cite{gatheral2011volatility}. We emphasize that using prices originating from a calibrated Heston model does not mean that we assume that the underlying security follows the dynamics of the Heston model, and the hedge itself remains model-free; however, in the absence of historical OTC prices, the empirical evaluation of the trading rule necessarily relies on surrogate prices generated from a calibrated model.}

\begin{figure}[htb!]
    \centering
    \includegraphics[width=0.45\linewidth]{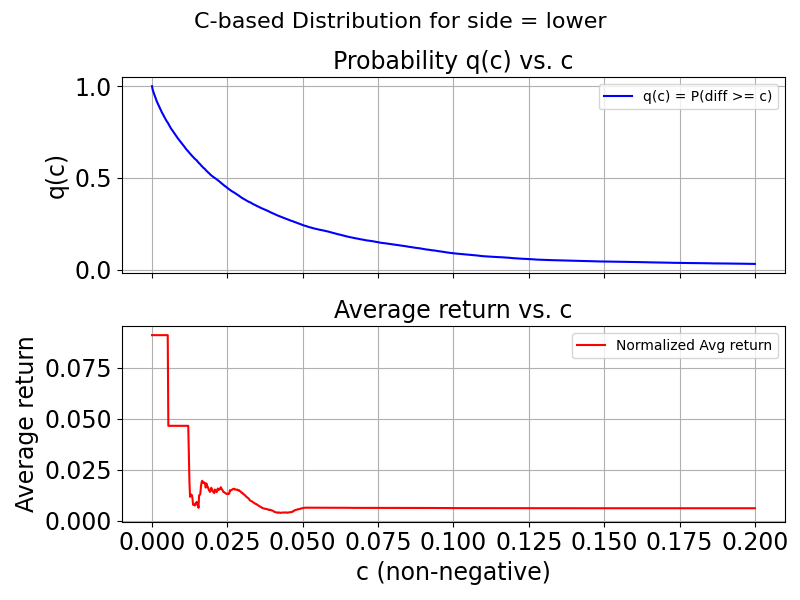}
        \includegraphics[width=0.45\linewidth]{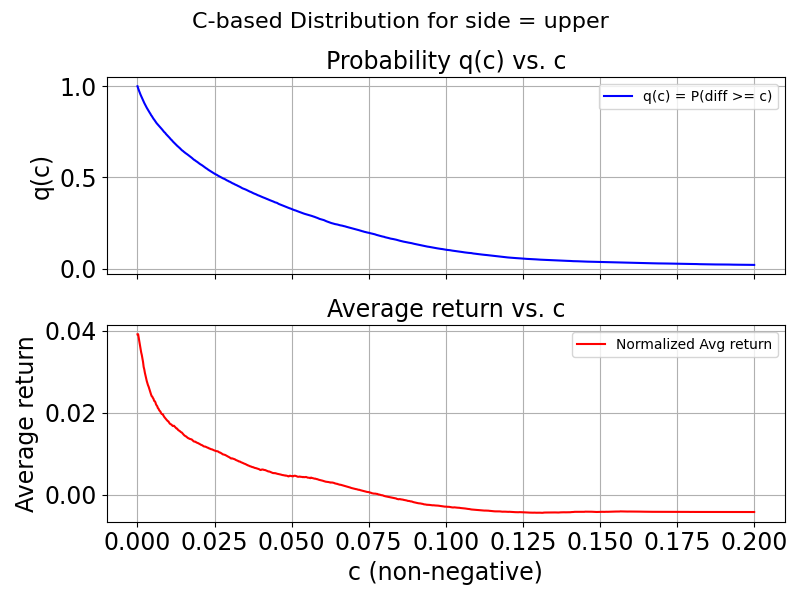}
    \caption{\textcolor{blue}{} Left: The graphs show $\underline{q}(c)$ and the average returns of $\frac{\Phi(S)-\underline{u_{d,(\theta_{i,j}),(H_i)}}(S)-\underline{C}}{S_0}$ conditional on $\frac{\mathfrak{M}(\Phi)-\underline{\operatorname{P}}(\Phi)}{S_0} \leq c$ in dependence of $c$, cost $\underline{C}=\mathfrak{M}(\Phi)-\underline{\operatorname{P}}(\Phi)$, evaluated on training data. \\ Right: The graphs show $\overline{q}(c)$ and the average returns of  $\frac{\overline{u_{d,(\theta_{i,j}),(H_i)}}(S)-\Phi(S)-\overline{C}}{S_0}$ conditional on $\frac{\overline{\operatorname{P}}(\Phi)-\mathfrak{M}(\Phi)}{S_0} \leq c$ in dependence of $c$, cost $\overline{C}=\overline{\operatorname{P}}(\Phi)-\mathfrak{M}(\Phi)$, evaluated on training data.}
    \label{fig:lowerBoundqrPlot}
\end{figure}

Figure~\ref{fig:lowerBoundqrPlot} illustrates the dependence of $\underline{q}(c)$ and $\overline{q}(c)$ on $c$ as well as the average returns of  $\frac{\Phi(S)-\underline{u_{d,(\theta_{i,j}),(H_i)}}(S)-\underline{C}}{S_0}$ conditional on ${\mathfrak{M}(\Phi)-\underline{\operatorname{P}}(\Phi)}\leq c\cdot S_0$, i.e., conditional on a trade being triggered as the market price is close enough to the lower model-free price bound, where we denote the costs of this strategy {(i.e., the costs of buying the option and selling the maximal sub-hedge)} by $\underline{C}:=\mathfrak{M}(\Phi)-\underline{\operatorname{P}}(\Phi)$. Analogously, we study $\frac{\overline{u_{d,(\theta_{i,j}),(H_i)}}(S) -\Phi(S)-\overline{C}}{S_0}$ conditional on ${\overline{\operatorname{P}}(\Phi)-\mathfrak{M}(\Phi)} \leq c\cdot S_0$ corresponding to the situation that a trade is triggered since the market price is close enough to the upper bound, where $\overline{C}:=\overline{\operatorname{P}}(\Phi)-\mathfrak{M}(\Phi)$ { denotes the costs of this strategy, i.e., the costs of selling the option and buying the minimal super-hedge}. We evaluate this on the training data. The two subplots at the bottom of Figure~\ref{fig:lowerBoundqrPlot}  show how the average net profit varies when gradually increasing $c$. Note that in this consideration the costs of the strategy are taken into account which explains the observation that with an increasing $c$  the average profit  decreases due to the increasing costs of the strategy.

In Table \ref{tab:upper} and Table \ref{tab:lower} we illustrate the results of the suggested strategy evaluated on training data, where in contrast to Figure~\ref{fig:lowerBoundqrPlot} we study annualized net profits per trade (no division by the spot value). By construction, both $\overline{q}(c)$ and $\underline{q}(c)$ decrease when $c$ increases and hence, combined with the increased costs of the strategy, the average return of the respective strategies decreases. Naturally, more strategies fulfil the selection criterion \eqref{eq:min_prices} with increasing $c$, implying that more trades are executed. An increasing $c$ also comes with a decreasing risk-adjusted return (in terms of Sortino ratio and Sharpe ratio). Also note that, since the strategies are almost not prone to losses (up to losing the initial investment), the variance is mainly driven by upward movements as being reflected in very high Sortino ratios. 

\begin{table}[htbp]
\caption{Train period performance results. Only trades triggered by $\overline{\operatorname{P}}(\Phi)-\mathfrak{M}(\Phi)\leq c \cdot S_0,$ are considered.}
\label{tab:upper}
\centering
\begin{tabularx}{\textwidth}{@{}l *{7}{>{\centering\arraybackslash}X}@{}}
\toprule
\textbf{$\overline{q}(c)$} & \textbf{$0.99$} & \textbf{$0.95$} & \textbf{$0.90$} & \textbf{$0.80$} & \textbf{$0.70$} & \textbf{$0.60$} & \textbf{$0.50$} \\
\midrule
\textbf{c} 
& {0.0002} & {0.0011} & {0.0025} & {0.0061} & {0.0113} & {0.0182} & {0.0272} \\

\textbf{Number of Trades} 
& {590} & {694} & {887} & {1456} & {2241} & {3220} & {4493} \\

\textbf{Avg return p.a.} 
& {1378.97} & {1280.04} & {1098.92} & {810.59} & {568.78} & {404.07} & {295.95} \\

\textbf{Std dev. p.a.} 
& {135.98} & {128.54} & {117.32} & {99.88} & {87.30} & {76.24} & {66.21} \\

\textbf{Down dev. p.a.} 
&  & {2.79} & {6.07} & {13.87} & {24.98} & {23.07} & {19.13} \\

\textbf{Sharpe ratio} 
& {10.14} & {9.96} & {9.37} & {8.12} & {6.52} & {5.30} & {4.47} \\

\textbf{Sortino ratio} 
&  & {458.99} & {181.03} & {58.46} & {22.77} & {17.51} & {15.47} \\

\textbf{Max. Loss} 
& {0.05} & {1.50} & {3.16} & {11.26} & {19.52} & {31.99} & {46.08} \\

\textbf{Max. Profit} 
& {388.72} & {388.72} & {388.72} & {388.72} & {388.72} & {388.72} & {388.72} \\

\bottomrule
\end{tabularx}
\end{table}

We observe in Table~\ref{tab:lower} that even for relatively large levels of $c$ only a small number of trades are executed because prices of the calibrated Heston model are much closer to the model-free upper bound, emphasizing the importance of including both lower bound and upper bound into the selection criterion \eqref{eq:min_prices}.

\begin{table}[htbp]
\caption{Train period performance results. Only trades triggered by $\mathfrak{M}(\Phi)-\underline{\operatorname{P}}(\Phi)\leq c \cdot S_0$ are considered.}
\label{tab:lower}
\centering
\begin{tabularx}{\textwidth}{@{}l *{7}{>{\centering\arraybackslash}X}@{}}
\toprule
\textbf{$\underline{q}(c)$} & \textbf{ $0.99$} & \textbf{$0.95$} & \textbf{$0.90$} & \textbf{$0.80$} & \textbf{$0.70$} & \textbf{$0.60$} & \textbf{$0.50$} \\
\midrule
\textbf{c} & {0.0000} & {0.0009} & {0.0021} & {0.0053} & {0.0093} & {0.0146} & {0.0208} \\
\textbf{Number of Trades} & {1} & {1} & {1} & {1} & {2} & {19} & {317} \\
\textbf{Avg return p.a.} & {5.05} & {5.05} & {5.05} & {5.05} & {3.07} & {1.27} & {28.70} \\
\textbf{Std dev. p.a.} &  &  &  &  & {0.18} & {0.18} & {89.36} \\
\textbf{Down dev. p.a.} &  &  &  &  &  & {0.02} & {57.52} \\
\textbf{Sharpe ratio} &  &  &  &  & {17.38} & {7.22} & {0.32} \\
\textbf{Sortino ratio} &  &  &  &  &  & {72.66} & {0.50} \\
\textbf{Max. Loss} & & & & & & {0.94} & {35.33} \\
\textbf{Max. Profit} & {9.97} & {9.97} & {9.97} & {9.97} & {9.97} & {9.97} & {181.30} \\

\bottomrule
\end{tabularx}
\end{table}

Next, we evaluate the performance of the strategy for different choices of $c$ on testing data and illustrate the results in Table~\ref{tab:testPriod}. The results confirm the profitability of the strategy, and they indicate that, as expected, in practice smaller levels of $c$ lead to higher risk-adjusted returns which however comes at the expense of trading less frequently, and therefore a restricted scalability of the strategy, which might in the extreme case even lead to no trades being triggered at all when choosing $c$ too small in an arbitrage-free market. 

{Next, we consider an alternative position-sizing rule according to which we invest at all times, and we let the investment amount depend on the distance to the nearest price bound: the closer the market price is to a bound (i.e., the smaller $c$), the larger the position we take. Specifically, we scale the investment by 
\begin{equation}\label{eq:scale_factor}
\exp\left(-\beta\cdot\left(\frac{\min\{\mathfrak{M}(\Phi)-\underline{\operatorname{P}}(\Phi), \overline{\operatorname{P}}(\Phi)-\mathfrak{M}(\Phi)\}}{S_0}\right)\right).
\end{equation}
Here, the term inside the minimum is the market price’s distance to the nearest no-arbitrage bound, and the parameter $\beta > 0$ controls how aggressively the investment is reduced as this distance increases: a larger $\beta$ leads to a sharper decline of the scaling factor from \eqref{eq:scale_factor} (i.e., investments tend to be concentrated on observations closer to the bounds). Results in the testing period can be found in Table~\ref{tab:testPriod_ex}. Note that the strategy trades at all times; however, by construction the position size becomes almost negligible when market prices are far from the price bounds.}
{
The optimal Sharpe ratio is attained for $\beta \approx 50$, whereas for the Sortino ratio, in line with the results from the first investment strategy, it seems to be optimal to choose $\beta$ as large as possible (meaning we prioritize strongly to invest  more for prices being close to the price bound). For $\beta =100$ we report a Sortino ratio of $7.29$.
}

\begin{table}[htbp]
\caption{Test period performance}
\label{tab:testPriod}
\centering
\begin{tabularx}{\textwidth}{@{}l *{10}{>{\centering\arraybackslash}X}@{}}
\toprule
\textbf{$c$} & \textbf{0\%} & \textbf{1\%} & \textbf{2.5\%} & \textbf{5\%} & \textbf{10\%}\\
\midrule
\textbf{$\overline{q}(c)$} 
& {1.00} & {0.89} & {0.67} & {0.34} & {0.06} \\
\textbf{$\underline{q}(c)$} 
& {1.00} & {1.00} & {0.91} & {0.00} & {0.00} \\
\textbf{Number of Trades} 
& {0} & {1607} & {6326} & {24962} & {29193} \\
\textbf{Avg return p.a.} 
&  & {630.71} & {515.06} & {386.22} & {353.90} \\
\textbf{Std dev. p.a.} 
&  & {298.68} & {304.28} & {369.80} & {343.07} \\
\textbf{Down dev. p.a.} 
&  & {42.35} & {54.33} & {74.60} & {68.30} \\
\textbf{Sharpe ratio} 
&  & {2.11} & {1.69} & {1.04} & {1.03} \\
\textbf{Sortino ratio} 
&  & {14.89} & {9.48} & {5.18} & {5.18} \\
\textbf{Max. Loss} 
&  & {40.53} & {111.15} & {242.57} & {303.75} \\
\textbf{Max. Profit} 
&  & {1341.05} & {2565.04} & {2565.04} & {3040.10} \\
\bottomrule
\end{tabularx}
\end{table}

{
\begin{table}[htbp]
\begingroup\color{black}
\caption{Test period performance with exponential investment}
\label{tab:testPriod_ex}
\centering
\begin{tabularx}{\textwidth}{@{}l *{5}{>{\centering\arraybackslash}X}@{}}
\toprule
\textbf{$\beta$} & \textbf{100} & \textbf{50} & \textbf{10} & \textbf{5} & \textbf{1} \\
\midrule
\textbf{Number of Trades} & 30162 & 30162 & 30162 & 30162 & 30162 \\
\textbf{Avg return p.a.} & 42.08 & 94.56 & 253.58 & 294.25 & 333.02 \\
\textbf{Std dev. p.a.} & 41.12 & 84.58 & 241.20 & 284.47 & 326.07 \\
\textbf{Down dev. p.a.} & 5.78 & 14.61 & 47.49 & 56.28 & 64.68 \\
\textbf{Sharpe ratio} & 1.02 & 1.12 & 1.05 & 1.03 & 1.02 \\
\textbf{Sortino ratio} & 7.29 & 6.47 & 5.34 & 5.23 & 5.15 \\
\textbf{Max. Loss} & 29.56 & 43.03 & 156.25 & 194.69 & 276.84 \\
\textbf{Max. Profit} & 939.04 & 1118.76 & 2080.93 & 2310.34 & 2863.32 \\
\bottomrule
\end{tabularx}
\endgroup
\end{table}
}

The substantially higher Sharpe and Sortino ratios in the training period should be interpreted with caution. They are in-sample quantities and are affected by the calibration of the empirical gap distribution, the selection of the threshold $c$, and the highly asymmetric payoff profile of the model-free strategy. The lower, but still positive, out-of-sample Sharpe ratios in Tables~\ref{tab:testPriod} and \ref{tab:testPriod_ex} indicate that part of the in-sample performance does not carry over to the test period. We suspect this gap between train and test performance is mainly due to a distribution shift between train and test set, in particular a change in $\overline{q}(c)$ and $\underline{q}(c)$.  Hence, the reported backtest should be viewed as evidence that the mechanism can generate economically meaningful opportunities under the stated assumptions, rather than as a definitive assessment of implementable trading performance.

\section{Conclusion}

In our study we have investigated the quality of model-independent super-hedging (and sub-hedging) strategies using data of liquidly traded options of constituents of the S\&P~500 index.
Our study reveals that  model-free pricing and hedging approaches lead, compared with model-based hedges based on Black–Scholes, local volatility or Heston models, (see Section~\ref{sec:model_based}) to competitive hedging results even though, by construction as worst-case approaches they are known to be conservative. 

Our quantification of the distribution of hedging gaps of model-free hedging strategies based on real data of the S\&P~500 index across moneyness and  different maturities  shows that these model-free hedges often lead to payoffs closely aligned with actual realized payoffs of the underlying derivatives. Building on these findings, we present in Section~\ref{sec:tradingstrat} a novel highly profitable low-risk trading strategy that exploits the empirical tightness of model-free bounds by trading in the model-free hedging whenever the market price is close enough to the theoretical price. 

Our findings indicate that model-free methods, traditionally viewed as overly conservative, can in fact yield efficient and statistically robust trading strategies given the underlying option data is sufficiently liquid and hence informative enough. This finding opens the door to broader practical adoption of robust approaches in the finance industry.

{
The approach presented in this paper can be easily extended to other financial instruments such as options on variance or volatility indices, e.g., VIX options or variance swap options. They provide a natural and promising setting for further market-based empirical tests of our framework, in particular, since these instruments are actively traded and therefore allow one to assess the performance of model-free hedging strategies using directly observed market prices rather than synthetically constructed ones.

To implement the strategy, the static component of the hedging strategy can be calibrated to observed option prices on volatility indices or variance swaps, while the dynamic component can be formulated with respect to the corresponding futures or index levels. This illustrates that our approach is not confined to synthetic examples but extends naturally to realistic market environments with liquidly traded derivatives. This setting therefore provides a natural bridge between robust, model-free hedging theory and empirically observable volatility markets.} 
\section*{Acknowledgements}
J. Sester gratefully acknowledges support by the NUS Start-Up Grant \emph{Tackling model
uncertainty in Finance with machine learning}, by the MOE AcRF Tier 1 Grant 	 25-0428-P0001, and by the MOE AcRF Tier 2 Grant 	
T2EP20225-0030.
\bibliographystyle{ecta} 
\bibliography{literature}

\appendix 

\section{Numerical Algorithm}
In this section we provide with Algorithm~\ref{alg:cutting_plane} a detailed description of the numerical algorithm used to compute model-free price bounds.

\begin{algorithm}[h!]
\SetAlgoLined
\SetKwInOut{Input}{Input}
\SetKwInOut{Output}{Output}

\Input{Number of traded call and put options $M_1$; bid and ask prices of traded call options; interest rate $r$; Payoff function $\Phi$; Tolerance level $\operatorname{TOL}$; Large grid $\G \subset \R^2$ with number of meshes $N^2$; Initial sub-grid $\G_0 \subset \G$; 
}
\For{$k = 0,1,\cdots$}{
Find feasible parameters $x^{(k)} := \left(d^{(k)},(\theta_{i,j}^{(k)})_{\substack{i=1,2\\ j =1,\dots,M_1 }},H_0^{(k)},({H_1^\ell}^{(k)})_{\ell=1,\dots,N}\right)$ that minimize \eqref{eq:super_replication_grid} on the grid $\G_k$, i.e.,
\[
\widetilde{\overline{\operatorname{P}}}(\Phi,\G_k) = d^{(k)}+\sum_{i=1}^2\sum_{j=1}^{M_i}({\theta_{i,j}^{\rm call, +}}^{(k)}h_{i,j}^{\rm call, +}-{\theta_{i,j}^{\rm call, -}}^{(k)}h_{i,j}^{\rm call, -})
\]
with $u_{x^{(k)}}(S) \geq \Phi(S) \text{ for all } S \in \G_k$ for $u_{x^{(k)}}$ as defined in \eqref{strategy_transaction_costs};

Compute the constraint violation on $\G$ given by $\delta:= \min_{S \in \G} \{ u_{x^{(k)}}(S) -\Phi(S)\}$;\\
\lIf{$\delta>-\operatorname{TOL}$}{STOP}
\Else{Define $\G_{k+1} := \G_k \cup \{ S \in \G~|~ u_{x^{(k)}}(S) -\Phi(S) <\delta+\frac{1}{k+1}\} ;$ }}
\Output{Parameters $d^{(k)},(\theta_{i,j}^{(k)})_{\substack{i=1,2\\ j =1,\dots,M_1 }},H_0^{(k)},({H_1^\ell}^{(k)})_{\ell=1,\dots,N}$;}

 \caption{Cutting plane algorithm to compute super-hedging strategies}\label{alg:cutting_plane}
\end{algorithm}

\section{Hyperparameter - Selection}\label{sec:data_hyper}

\subsection{Parameter optimization}

Applying Algorithm~\ref{alg:cutting_plane} requires selecting hyperparameters that are used as input to the algorithm. To find best-working hyperparameters and understand its influence on the outcomes, we conducted a systematic exploration of the involved hyperparameters, involving the parameters tolerance ($\rm TOL$), initial grid size ($n$), and large grid size ($N$).

\subsubsection{Superhedging gap and cutting-plane algorithm}

For our analysis, Algorithm~\ref{alg:cutting_plane} was applied to compute the gap between the superhedging value and the actual hedge performance on a subset of the dataset described in Section~\ref{sec:Data_and_tech}.

\begin{figure}[h]
        \centering
        \includegraphics[width=0.45\textwidth]{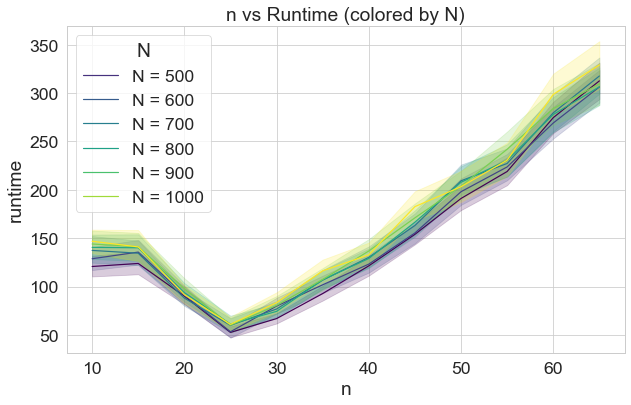} 
            \includegraphics[width=0.45\textwidth]{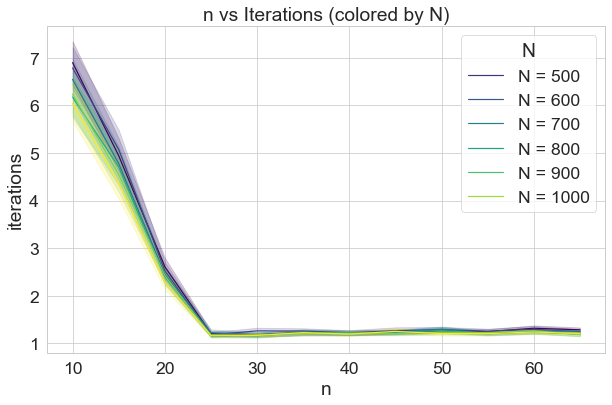} 
        \caption{Left: The plot depicts how different grid sizes influence the computational time, Right: The plot illustrates the interaction between grid sizes ($N$) and the number of iterations ($n$) required for convergence (reaching the tolerance level). }        \label{fig:runtime}
    \end{figure}

\begin{figure}[h]
        \centering
        \includegraphics[width=0.45\textwidth]{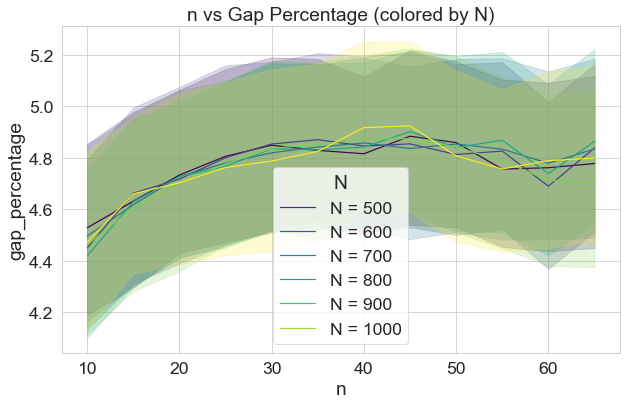} 
             \includegraphics[width=0.45\textwidth]{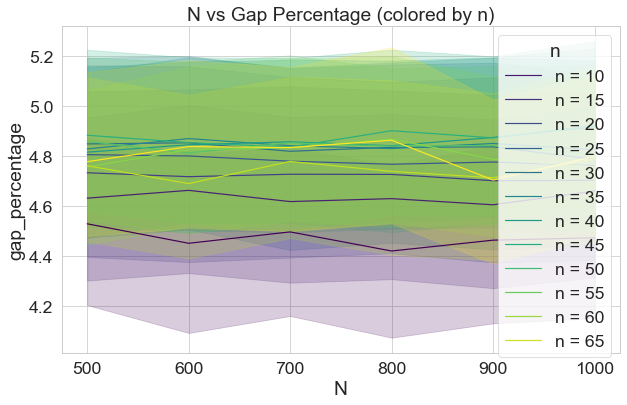} 
        \caption{The plots illustrate the impact of grid sizes ($N$) and the number of iterations ($n$) on the gap between superhedging and actual payoff of the derivative.}        \label{fig:Ngap}
    \end{figure}

To understand the effects of the parameters on the algorithm's performance, we visualize in Figures~\ref{fig:runtime} and \ref{fig:Ngap} the effects of varying the parameters on the total time taken for the cutting plane algorithm to converge (\textbf{Runtime}), the number of iterations required for convergence (\textbf{Iteration count}) and on the relative size of the difference between the superhedging value and actual hedge (\textbf{Gap percentage}).

These visualizations helped us identify optimal parameter combinations that minimize both the gap and runtime while ensuring sufficient convergence. Details of the used hyperparameters can be found under  \href{https://github.com/QIYIHAN/Empirical-Price-Bounds}{https://github.com/QIYIHAN/Empirical-Price-Bounds}.

\section{Distribution of the payoff gaps}\label{sec:hist}
\begin{figure}[h!]
    \centering
    \includegraphics[width=0.9\linewidth]{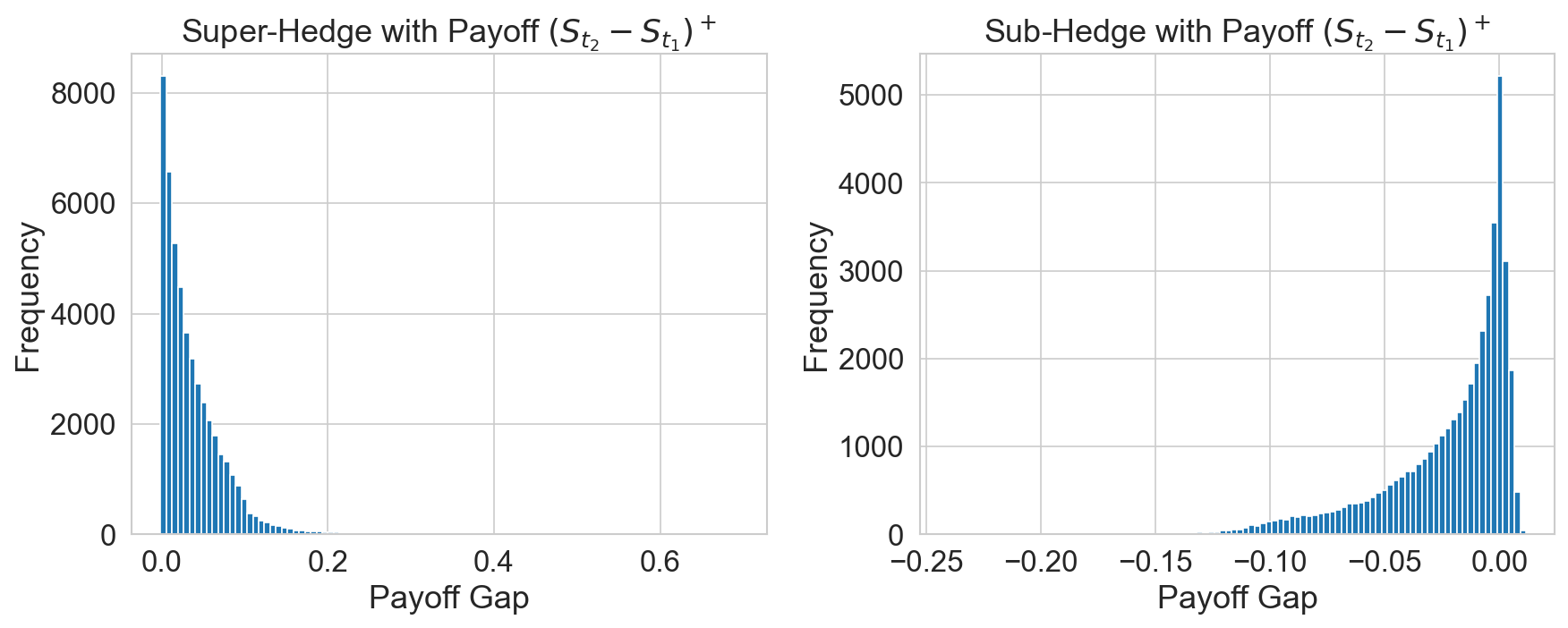}
    \caption{\textcolor{blue}{}The histogram of the hedging gap with payoff function $(S_{t_2}-S_{t_1})^+$.}
    \label{fig:payoff1_hist}
\end{figure}

\begin{figure}[h!]
    \centering
    \includegraphics[width=0.9\linewidth]{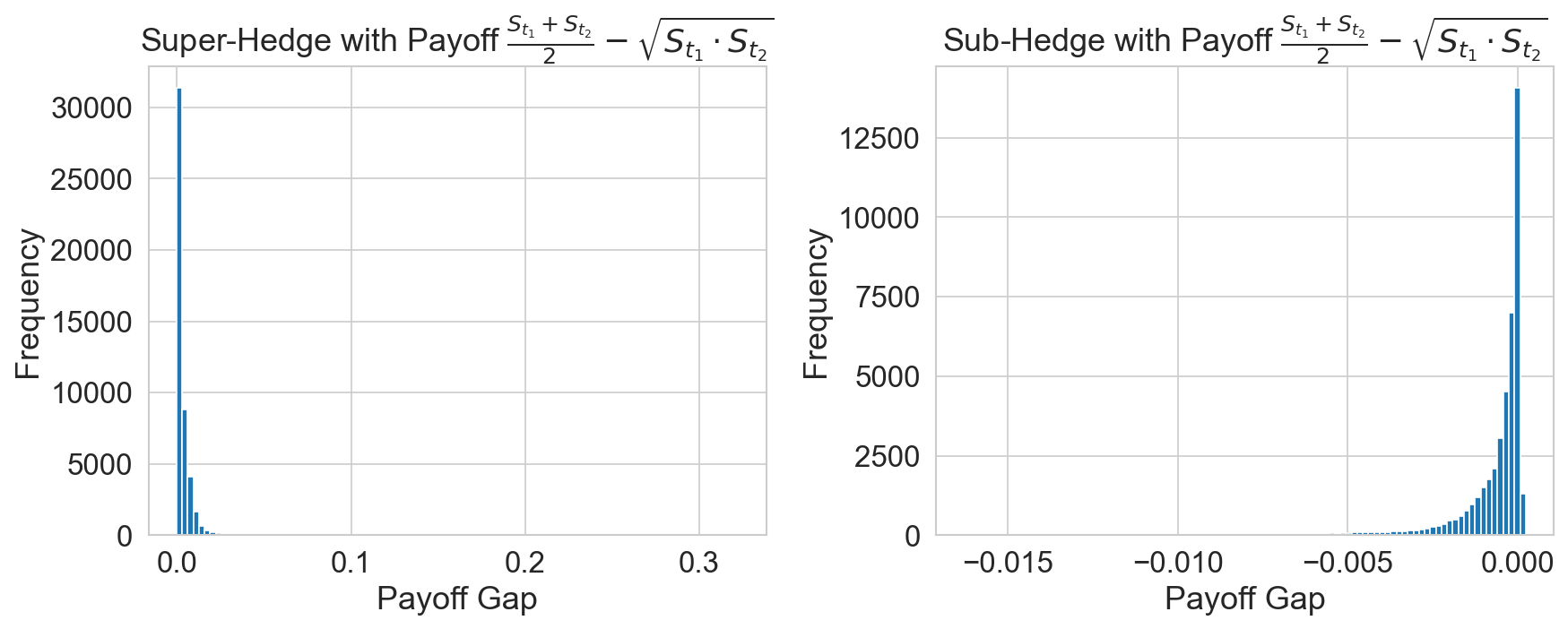}
    \caption{\textcolor{blue}{}The histogram of the hedging gap  with payoff function $\frac{S_{t_1}+S_{t_2}}{2}-\sqrt{S_{t_1}\cdot S_{t_2}}$.}
    \label{fig:payoff7_hist}
\end{figure}

\begin{figure}[h!]
    \centering
    \includegraphics[width=0.9\linewidth]{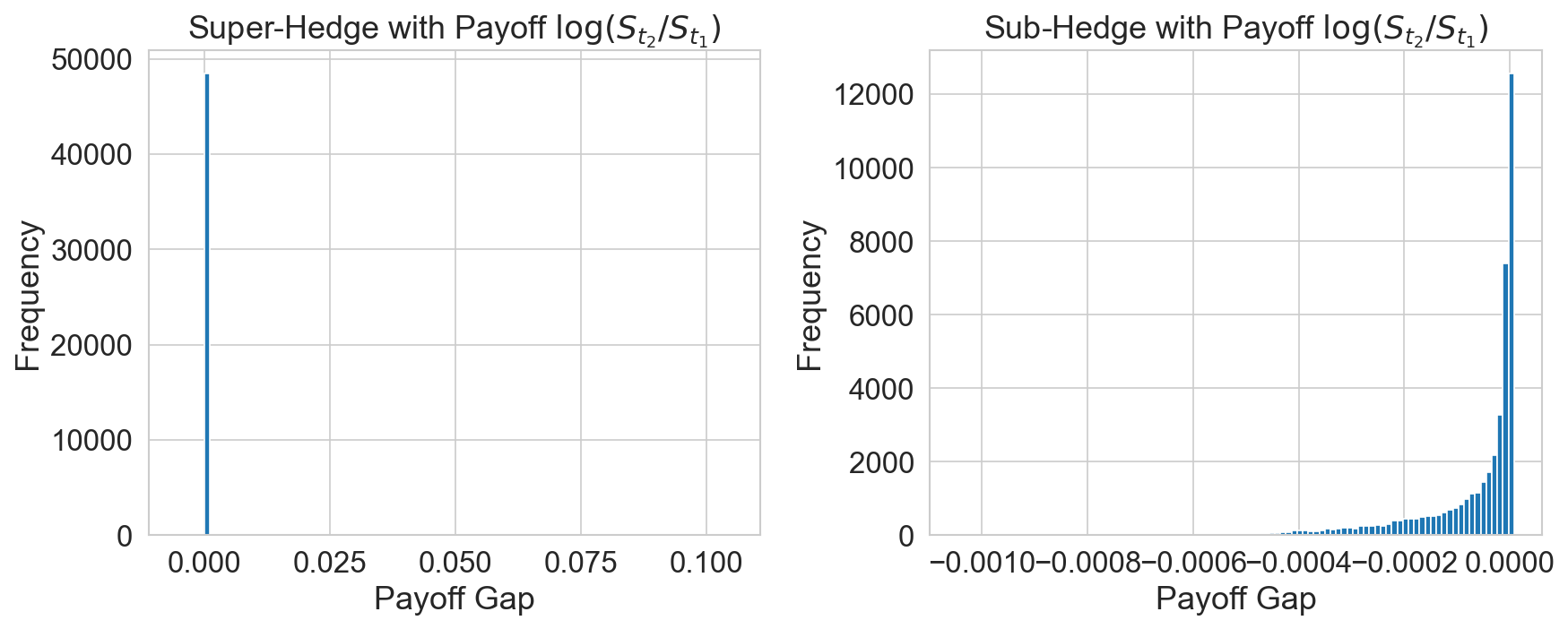}
    \caption{\textcolor{blue}{}The histogram of the hedging gap  with payoff function $\log(S_{t_2}/S_{t_1})$.}
    \label{fig:payoff3_hist}
\end{figure}

\begin{figure}[h!]
    \centering
    \includegraphics[width=0.9\linewidth]{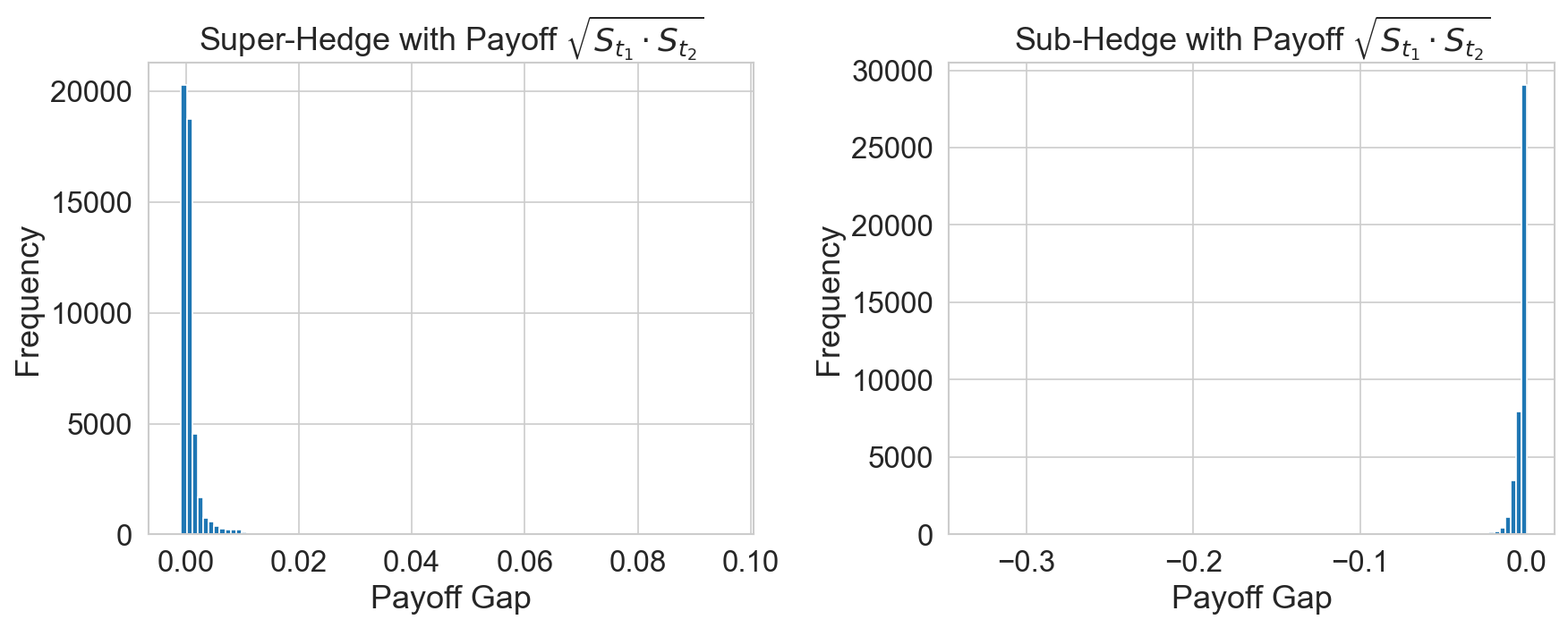}
    \caption{\textcolor{blue}{}The histogram of the hedging gap  with payoff function $\sqrt{S_{t_1}\cdot S_{t_2}}$.}
    \label{fig:payoff4_hist}
\end{figure}

Here, in Figure~\ref{fig:payoff1_hist}--\ref{fig:payoff4_hist} we provide histograms illustrating the distributions of the payoff gaps of the experiments from Section~\ref{sec:superhedge_forward}. 

\section{Benchmark Models}\label{sec:models}
In this section we briefly describe the benchmark models used in Section~\ref{sec:model_based} and the method to compute prices of forward start option and, based on this, its delta sensitivities. For all of the models, we approximate the option delta as $\Delta(S_t,t):= \frac{{\rm P}\left(S_t\cdot (1+\alpha),t\right))-{\rm P}(S_t,t)}{\alpha S_t}$ on a daily basis, where we use $\alpha= 0.001$, and where ${\rm P}(S_t,t)$ denotes the option price in the respective model at time $t$ when the underlying security attains value $S_t$. 
\subsubsection*{Black--Scholes model}
The Black--Scholes model (\cite{black1973pricing}) models the evolution of the asset value $S_t$ by the stochastic differential equation
\begin{align*}
    \D S_t &= r S_t \D t + \sigma S_t \D W_t
\end{align*}
for some $r\in \R$, $\sigma >0$ and some standard Brownian motion $(W_t)_{t \geq 0}$. We use the same interest rate $r$ as in the model-free considerations, described in Section~\ref{sec:ir} as well as the Black--Scholes implied volatility $\sigma^{\rm BS}$, i.e., the volatility $\sigma^{\rm BS}$ that makes the call option market prices match the model-based call option prices. For the Black--Scholes model, the option price $P(S,t)$ is calculated as the analytical price of the forward-start option price formula with $k=1$ in \cite[p. 231]{musiela1997martingale}.
\subsubsection*{Heston model}\label{subsubsec:heston}
As a second benchmark model we consider the stochastic volatility model introduced by \cite{heston1993closed} where the stock price $(S_t)_{t\geq 0}$ in the continuous-time \emph{Heston model} is given by the stochastic differential equations
\begin{align*}
    \D S_t &= r S_t \D t + \sqrt{\nu_t} S_t \D W_t^1, \\
    \D \nu_t &= \kappa (\theta - \nu_t) \D t + \sigma \sqrt{\nu_t} \D W_t^2,
\end{align*}
for $r$ being the interest rate, $\kappa$ the volatility mean-reversion speed, $\theta$ the volatility long term mean, $\sigma$ the diffusion parameter associated to the volatility, and where $W_t^1$ and $W_t^2$ are two correlated Brownian motions. The parameters of the Heston model are calibrated to the considered call option prices according to the methodology described in \cite{levendorskiui2012efficient}. 
 On each trading day, we use data of all option prices, stock prices and interest rates of this day, and calibrate Heston parameters for each underlying stock. To calibrate the Heston model parameters, we minimize the squared distance between observed market prices and model-implied prices obtained under the Heston dynamics. To enhance numerical stability and mitigate sensitivity to small perturbations in the input data, we additionally incorporate a Tikhonov-type regularization term (see, e.g., \cite{egger2005tikhonov}) that penalizes abrupt temporal changes in the parameter estimates. The resulting calibration problem is formulated as
$$
\min_{(\Theta_{t,j})_{j=1,\dots,m}}\{\sum_{i=1}^n\left \lVert \operatorname{P}(\Phi_i,\Theta_t) - \mathfrak{M}(\Phi_i) \right\rVert_2^2+\alpha\sum_{j=1}^m \left\lVert \Theta_{t,j} - \Theta_{t-1,j} \right\rVert_2^2\},
$$ where $\mathfrak{M}(\Phi_i)$, denotes the market price of the 
$i$-th option, and  $\operatorname{P}(\Phi_i,\Theta_t)$  is the corresponding model price under the Heston model with parameter vector $\Theta_t$ at time $t$. Here, $\Theta_t = (\Theta_{t,j})_{j=1,\dots,m}$,\  collects the $m=5$ Heston parameters at time $t$ and $\alpha>0$ controls the strength of regularization.
Once the model parameters have been calibrated, the price of a forward-start option, and, consequently, its sensitivities, can be computed in closed form. We refer to \cite{Kruse2005} for the explicit pricing formula.

\subsubsection*{Dupire's Local Volatility model}
Dupire's local volatility model (\cite{dupire1994pricing} and \cite{derman1994riding}) models the asset price $(S_t)_{t \geq 0}$ via
\begin{align*}
    \D S_t &= r S_t \D t + \sigma(t,S_t) S_t \D W_t
\end{align*}
for interest rate $r$ and Brownian motion $(W_t)_{t \geq 0}$ as well as
\begin{equation}\label{eq:locvol_dupire}
\sigma^2(T,K)=2\frac{\frac{\partial}{\partial T}c(T,K)+Kr\frac{\partial}{\partial K}c(T,K)}{K^2 \frac{\partial^2}{\partial K^2}c(T,K)}
\end{equation}
with $K$ and $T$ being strike and maturity, respectively of the considered call option with price $c(T,K)$. The derivatives in the right-hand-side of \eqref{eq:locvol_dupire} are approximated via finite differences. 

The delta sensitivity is then derived from a corresponding Black--Scholes model, considering the option price as a function of stock price $S$ with volatility $\sigma$, $P = P(S,\sigma(t,S))$. The option delta can then be calculated analytically as 
$$\frac{dP}{dS} = \frac{\partial P}{\partial S}\bigg|_{\sigma} + \frac{\partial P}{\partial \sigma}\bigg|_{S} \frac{\partial \sigma(T, K, S)}{\partial S}
= \Delta_{\text{BS}}(S, \sigma) + \nu_{\text{BS}}(S, \sigma) \frac{\partial \sigma(T, K)}{\partial S}$$
for $\Delta_{\text{BS}}$ and $\nu_{\text{BS}}$ denoting Delta and Vega, respectively in the Black--Scholes model.

\end{document}